\documentclass[aps,prd,groupedaddress,twocolumn,showpacs,nofootinbib,superscriptaddress]{revtex4-1}

\pdfoutput=1

\usepackage{graphicx,epsfig}
\usepackage{amssymb,amsmath}

\begin{document}

\title{Coupled Quintessence and the Halo Mass Function}


\author{Ewan R. M. Tarrant}
\email[]{ppxet@nottingham.ac.uk}
\affiliation{School of Physics and Astronomy, University of Nottingham, University Park, Nottingham, NG7 2RD, UK}

\author{Carsten van de Bruck}
\email[]{c.vandebruck@sheffield.ac.uk}
\affiliation{Department of Applied Mathematics, The University of Sheffield, Hounsfield Road, Sheffield, S3 7RH, UK}

\author{Edmund J. Copeland}
\email[]{ed.copeland@nottingham.ac.uk}
\affiliation{School of Physics and Astronomy, University of Nottingham, University Park, Nottingham, NG7 2RD, UK}

\author{Anne M. Green}
\email[]{anne.green@nottingham.ac.uk}
\affiliation{School of Physics and Astronomy, University of Nottingham, University Park, Nottingham, NG7 2RD, UK}


\date{\today}

\begin{abstract}
A sufficiently light scalar field slowly evolving in a potential can account for the dark energy that presently dominates the universe. This quintessence field is expected to couple directly to matter components,  unless some symmetry of a more fundamental theory protects or suppresses it. Such a coupling would leave distinctive signatures in the background expansion history of the universe and on cosmic structure formation, particularly at galaxy cluster scales. Using semi--analytic expressions for the CDM halo mass function, we make predictions for halo abundance in models where the quintessence scalar field is coupled to cold dark matter, for a variety of quintessence potentials. We evaluate the linearly extrapolated density contrast at the redshift of collapse using the spherical collapse model and we compare this result to the corresponding prediction obtained from the non--linear perturbation equations in the Newtonian limit. For all the models considered in this work, if there is a continuous flow of energy from the quintessence scalar field to the CDM component, then the predicted number of CDM haloes can only lie below that of $\Lambda$CDM, when each model shares the same cosmological parameters today. In the last stage of our analysis we perform a global MCMC fit to data to find the best fit values for the cosmological model parameters. We find that for some forms of the quintessence potential, coupled dark energy models can offer a viable alternative to $\Lambda$CDM in light of the recent detections of massive high--$z$ galaxy clusters, while other models of coupled quintessence predict a smaller number of massive clusters at high redshift compared to $\Lambda$CDM.
\end{abstract}


\maketitle


\section{Introduction}\label{intro}
 
During the last decade, a vast amount of cosmological data has been collected, ranging from measures of the Hubble constant ($H_{0}$)~\cite{HST} to precision measurements of the Cosmic Microwave Background (CMB)~\cite{WMAP7}, observations of Type Ia Supernovae (SNIa)~\cite{kowalski} and Baryon Acoustic Oscillations (BAO)~\cite{percival}. These datasets are consistent with the standard $\Lambda$CDM cosmological model, which consists of a cosmological constant ($\Lambda$) plus Cold Dark Matter (CDM). However  this model suffers from several short-comings, namely the \textit{fine--tuning} and \textit{coincidence} problems (see e.g. \cite{carroll} for a review).

These issues surrounding the cosmological constant have prompted investigation into alternative models~\cite{starobinsky, carroll, padmanabhan}, such as quintessence~\cite{wetterich, ratrapeebles, wetterich2}, whereby the Dark Energy (DE) component is dynamic and is identified with a light scalar field $\phi(t)$, slowly evolving in a potential $V(\phi)$ (for a comprehensive review see \cite{Copeland:2006wr}). Since this quintessence field must be sufficiently light to drive an accelerated expansion, we should expect it to couple to other matter forms in the universe, unless some symmetry of a more fundamental theory protects or suppresses it. Whilst a coupling to baryonic matter is tightly constrained by local gravity tests~\cite{carrollQrestWorld}, interactions within the dark sector have not been ruled out by observations. A number of previous studies examined the cosmological implications of interacting DE--CDM cosmologies (see e.g.~\cite{amendolaa, amendolab, pettorinoa,Brookfield} and references therein) while analysis of CMB and BAO have placed constraints on various coupled quintessence models~\cite{amendolaCamposRosenfeld, beanTrodden}. These studies have shown that dark sector interactions can leave distinctive signatures in the background expansion history of the universe and on cosmic structure formation~\cite{baldi0, baldib, mainini,sutter}, at large (galaxy cluster) scales in particular.

Intriguingly, a number of massive, high redshift clusters have recently been detected~\cite{bremer, mancone, muchovej} which should be extremely rare in the standard $\Lambda$CDM model. It is possible that coupled quintessence cosmologies (hereafter CQ) may alleviate these tensions by modifying the expansion history and perturbation growth~\cite{Baldi:2010td}. It is therefore important to understand structure formation in CQ models. 

In this paper we consider three different quintessence potentials and examine how the strength of the DE--CDM coupling (which we take to be constant) affects the abundance of CDM haloes. We present results for three commonly used analytic forms for the CDM halo mass function, Jenkins et al.~\cite{jenkins}, Press--Schechter~\cite{press} and Sheth--Tormen~\cite{sheth}. How the dark energy scalar field behaves in highly non--linear regimes has yet to be understood and properly modelled. Indeed, the evaluation of the linearly extrapolated density contrast at the redshift of collapse, $\delta_{\star}(z)$, in CQ cosmologies has recently been the subject of some debate~\cite{wintergerst}. To get a feeling for the uncertainty when modelling the non--linear behaviour of the scalar field, we consider two different methods for evaluating $\delta_{\star}(z)$, (i) the spherical collapse model applied to CQ, and (ii) the prediction obtained from the non--linear perturbation equations in the Newtonian limit (see Ref.~\cite{wintergerst}). We first assume that all the CQ models share the same cosmological parameters as $\Lambda$CDM at $z = 0$. Whilst this facilitates an easier comparison between the models, 
as emphasised by Ref.~\cite{jennings} for the case of uncoupled quintessence, it is unlikely that they are consistent with observational data. Therefore in the final stage of our analysis we perform a global fit using the \texttt{CosmoMC} package~\cite{cosmomc}, a Monte Carlo Markov Chain (MCMC) code, to find the best fit values for the cosmological parameters such that each model satisfies the CMB+BAO+SN1a+$H_{0}$ datasets. Similar analysis have been carried out for a large number of non--interacting dark energy models~\cite{basilakos1,basilakos2}.

In Section \ref{CQcosmos} we introduce the quintessence models under investigation. Here we study the evolution of the background and perturbed quantities, pointing out the distinctive features of each model.  We discuss the halo mass functions that we choose to work with in Section \ref{massfunction}, paying special attention to the linearly extrapolated density contrast at the redshift of collapse $\delta_{\star}(z)$. In Section \ref{results} we present our predictions for halo abundance including a global MCMC fit for each model and we conclude in Section \ref{conclusions}. \\


\section{Coupled quintessence cosmologies}\label{CQcosmos}

Whilst in any multi--component system the total stress--energy momentum tensor of the Universe must be conserved~\cite{kodamaSasaki},

\begin{equation}
	\nabla_{\mu}\sum_{i}T^{\mu}_{(i)\nu}=0 \,,
	\label{conserveTmunu}	
\end{equation}

\noindent the stress--energy momentum tensor for each individual species $i$ may in general however, not be conserved. We consider cosmologies comprised of radiation ($\gamma$), CDM ($c$), baryons ($b$) and the quintessence field ($\phi$). Neutrinos are assumed to be massless, forming a fraction of the radiation content. Then, subject to the above constraint, we may consider a dark sector coupling~\cite{amendolaa}

\begin{eqnarray}
	 \nabla_{\mu}T^{\mu}_{({\rm c})\nu}    &=&  Q_{({\rm c})\nu} \,,  \nonumber \\
	 \nabla_{\mu}T^{\mu}_{(\phi)\nu} &=& Q_{(\phi)\nu} \,,
	\label{TmunuCoupled}	
\end{eqnarray}

\noindent where the source terms $Q_{(i)\nu}$ represent the interaction. We adopt natural units and set $8\pi G_{\rm N}=M^{-2}_{\text{pl}}=1$. If the quintessence field is coupled only to CDM, the constraint Eq.~(\ref{conserveTmunu}) requires that $Q_{(\phi)\nu}=-Q_{({\rm c})\nu}$. Our set of cosmologies are described by a Lagrangian of the form

\begin{equation}
	\mathcal{L}=-\frac{1}{2}\partial^{\mu}\phi\partial_{\mu}\phi-V(\phi)+\mathcal{L}_{\text{\rm matter}}[\chi_i,\phi] \,,
	\label{quintLagrangian}	
\end{equation}

\noindent where $\chi_i$ are the matter fields (baryons, cold dark matter, neutrinos). Here, we consider a model in which only cold dark matter couples to the quintessence scalar field $\phi$ and in which the mass of the dark matter field is a function of $\phi$: $m=m(\phi)$. The choice $m(\phi)$ specifies the source term, which as is customary, we define as 

\begin{equation}
	Q_{({\rm c})\nu} = -\beta \rho_{{\rm c}}\partial_{\nu}\phi \,,
	\label{defQ}	
\end{equation}

\noindent with

\begin{equation}
	\beta \equiv \frac{\partial\,\ln{m(\phi)}}{\partial\phi} \,.
	\label{defBeta}	
\end{equation}

\noindent An immediate consequence of Eq.~(\ref{defBeta}) is that the mass of the CDM matter particles no longer scale as $a^{-3}$ and hence (with $\beta$ constant)

\begin{equation}
	\rho_{c}=\rho_{{\rm c},0}a^{-3}e^{\beta\phi} \,.
	\label{massCDMCQ}	
\end{equation}

\noindent A coupling of the form Eq.~(\ref{defBeta}) can arise for instance after a conformal transformation of a two--metric Brans--Dicke theory~\cite{damouretal} which leaves radiation and baryons uncoupled. Radiation remains uncoupled since its stress--energy momentum tensor is traceless, whilst a coupling to baryonic matter is tightly constrained by local gravity tests~\cite{carrollQrestWorld}. We will assume that the coupling $\beta$ in Eq.~(\ref{defBeta}) is constant throughout this work. The background and linear perturbation evolution for this set of coupled dark energy models has been widely studied for constant $\beta$,~\cite{wetterich2, amendolaa} as well as with regard to the effects on structure formation~\cite{baldi0,mainini,maccio}. More recently, N--body simulations involving time varying couplings $\beta(t)$ have revealed several interesting effects in the non--linear regime of structure formation~\cite{baldib}.

The zero--component of Eq.~(\ref{TmunuCoupled}) (where $Q_{({\rm c})0} = -\beta \rho_{{\rm c}}\dot{\phi}$) provides the conservation equations for the energy densities of each species. For $Q_{({\rm c})0}>0$  $(<0)$, the direction of energy transfer is from the CDM (quintessence field) to the quintessence field (CDM). The direction of energy exchange then depends upon the evolution of the dark energy scalar field $\phi$ and on the sign of the coupling, $\beta$. Once a non--zero coupling of the form Eq.~(\ref{defBeta}) is introduced, the evolution of the field is driven by the properties of the effective potential, $V_{{\rm eff}}=V(\phi)+\rho_{{\rm c}}(\phi)$, which may possess a minimum about which the field may undergo damped oscillation. Hence, the scalar field velocity $\dot{\phi}$ may switch sign, reversing the direction of energy exchange. Such sign changeable interactions have been studied in the past for models where the role of dark energy is played by a decaying vacuum energy~\cite{RongGen, HaoWei}.


\subsection{Quintessence potentials}\label{potentials}

\begin{figure*}[t]
	\begin{tabular}{cc}
		\includegraphics[width=8.5cm]{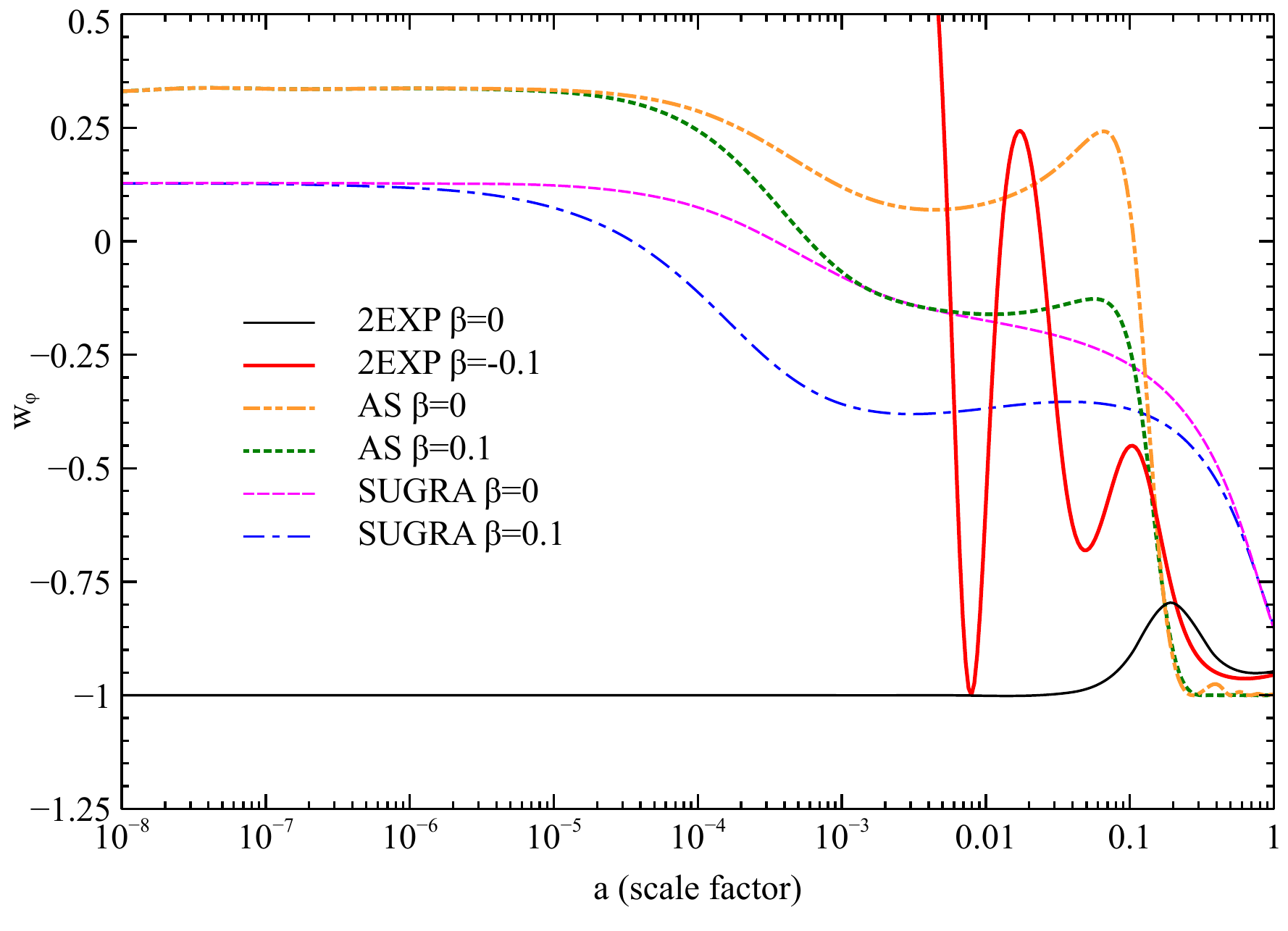} &
		\includegraphics[width=8.6cm]{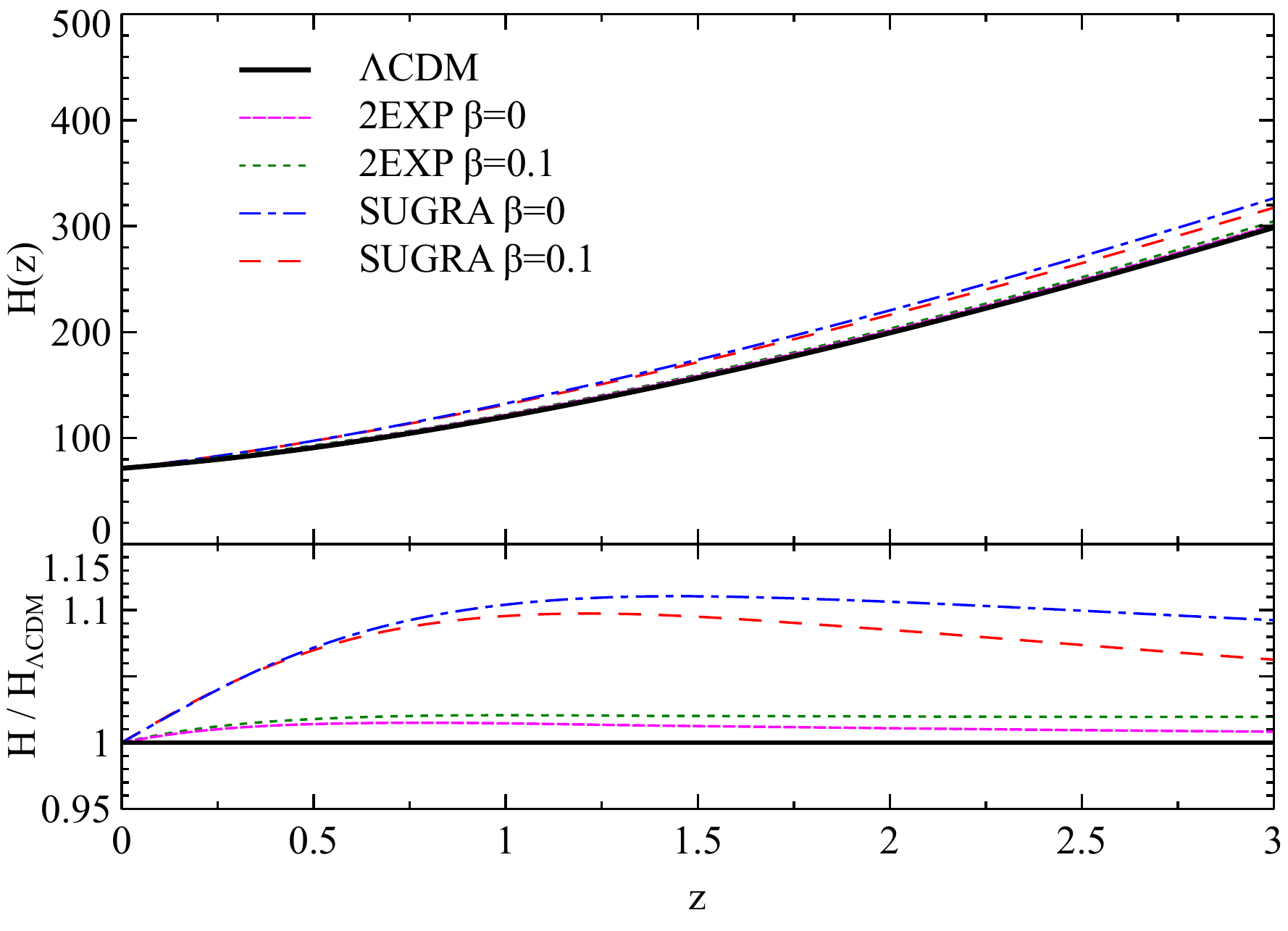} 
	\end{tabular}	
	\caption{\textit{Left panel}: The evolution of the quintessence field equation of state parameter $w_{\phi}=P_{\phi}/\rho_{\phi}$ for the three potentials for a variety of coupling strengths $\beta$. \textit{Right panel}: The redshift evolution of the Hubble parameter $H(z)$ in the 2EXP and SUGRA models. In the lower panel we compare to the evolution of $H(z)$ in $\Lambda$CDM by plotting the ratio. For the AS model, $H(z)$ is identical to $\Lambda$CDM over the redshift range of interest, for all coupling strengths considered in this study.}
	\label{fig:EoS}
\end{figure*}

In this study, we will be concerned with three different forms for $V(\phi)$ which we briefly discuss below. These potentials have the form $V=V_{0}f(\phi;\lambda_{n})$, where $V_{0}$ is a constant and $f(\phi;\lambda_{n})$ is a function of the field $\phi$, with the quintessence model fixed by choice of the self coupling parameters $\lambda_{n}$ and the strength of the DE--CDM coupling $\beta$. We numerically integrate the background equations of motion using a version of the \texttt{CAMB}~\cite{CAMB} computer software that we have modified to include a quintessence scalar field coupled to CDM. To facilitate an easier comparison between each model, the parameters $\lambda_{n}$ in the potential are kept fixed whilst $V_{0}$ and the CDM density $\rho_{c,0}$ are varied (for a given $\beta$) such that each cosmology evolves to the $\Lambda$CDM WMAP7 Maximum Likelihood (ML) cosmological parameter values~\cite{WMAP7}. The same initial conditions $\phi_{i}=10^{-4}$ and $\dot{\phi_{i}}=0$ for the field are assumed for each model.

The resulting cosmology may not be consistent with observational datasets and so in Section \ref{resultsB} we perform a global MCMC fit for each quintessence model to determine $\beta$ and the values of other cosmological parameters. \\


\textit{Double exponential (2EXP)} -- This potential is an example of a scaling solution and requires no fine tuning of the initial conditions~\cite{barreiro}. It is given by

\begin{equation}
	V(\phi)=V_{0}[e^{\lambda_{1}\phi}+e^{\lambda_{2}\phi}] \,.
	\label{2EXPpotential}	
\end{equation}

\begin{figure*}[t]
	\begin{tabular}{cc}
		\includegraphics[width=8.7cm]{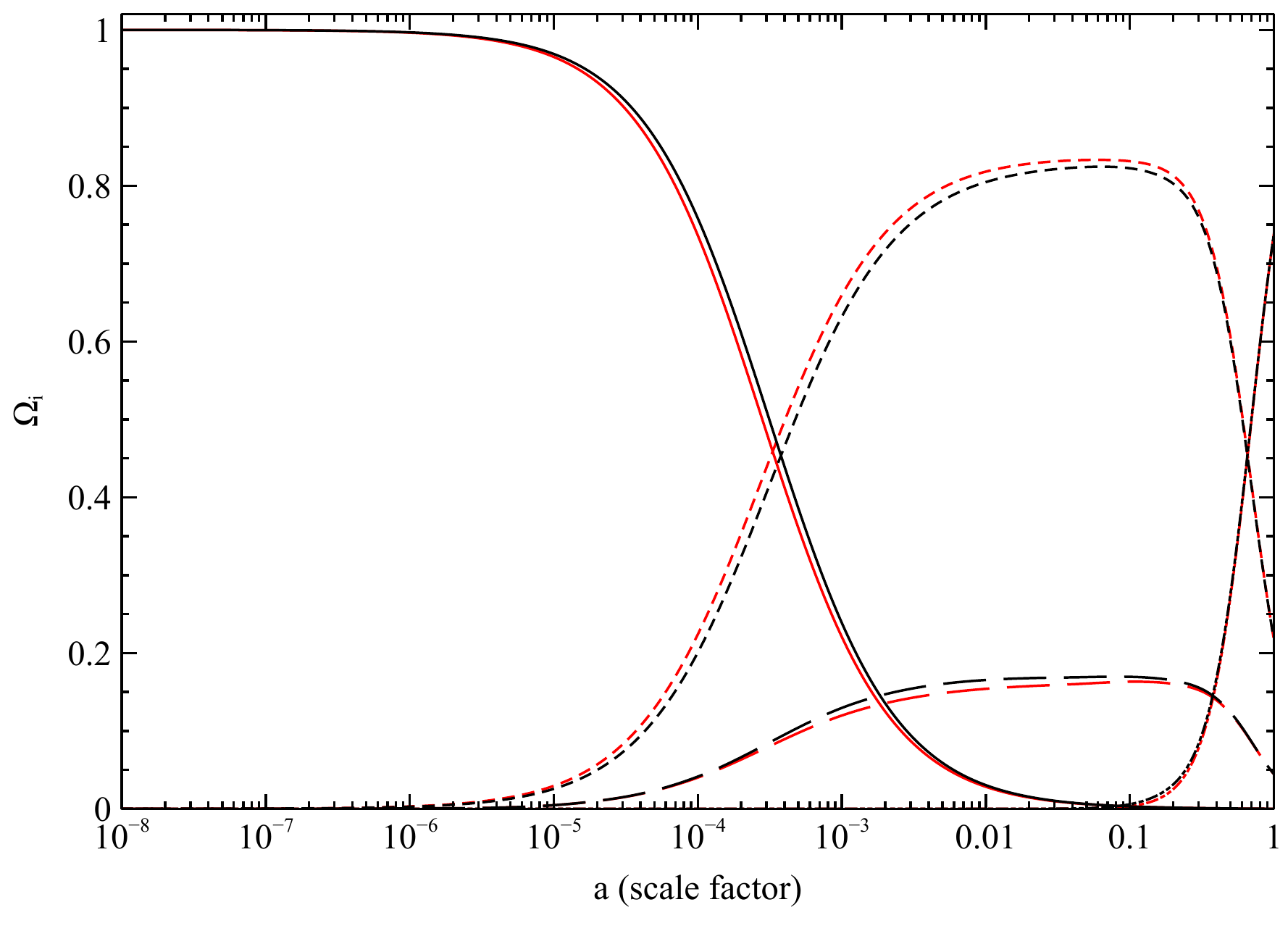} &
		\includegraphics[width=8.7cm]{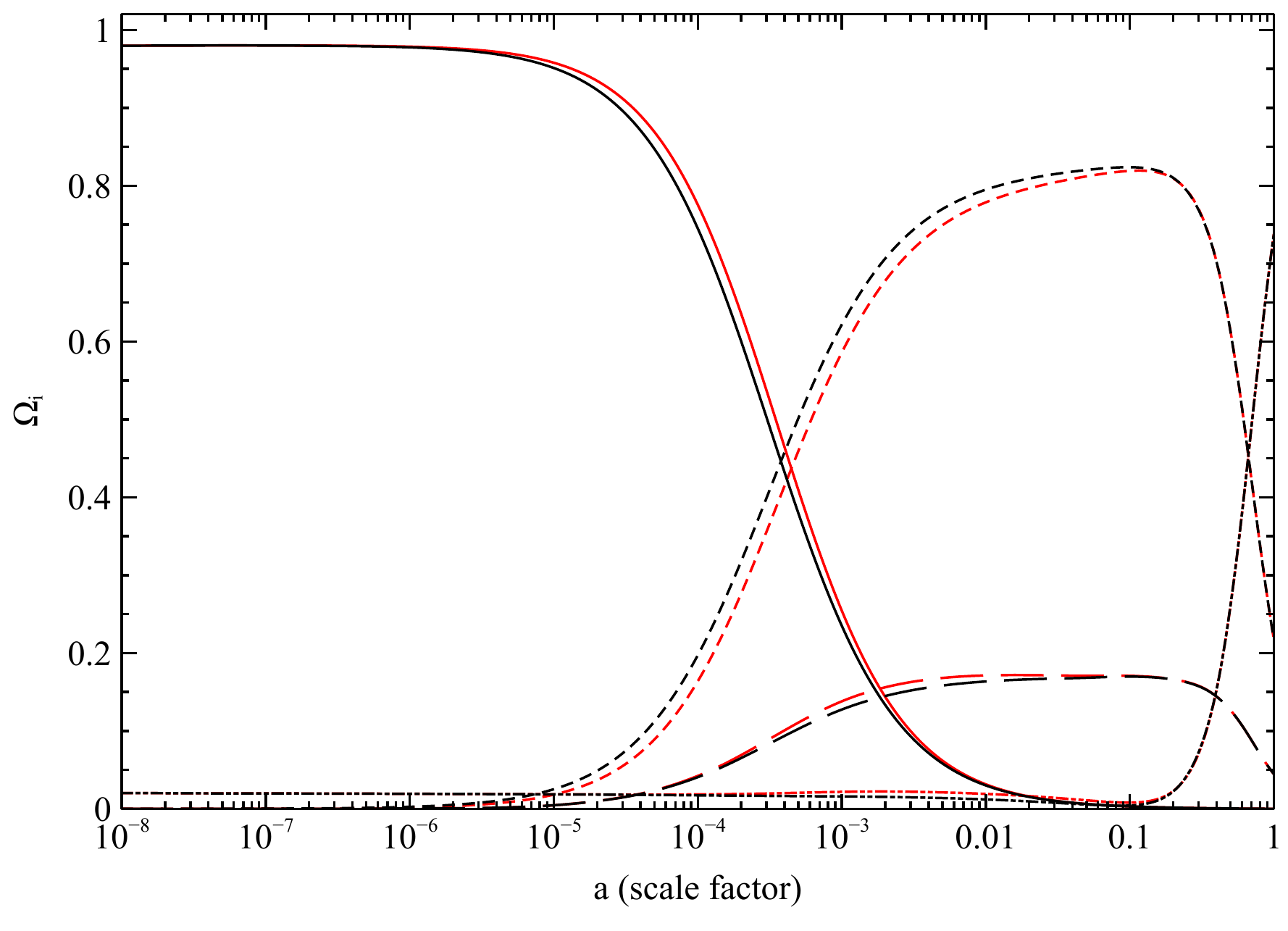}
	\end{tabular}
	\caption{The cosmological evolution of the energy density parameters $\Omega_{i}$ in the 2EXP (\textit{left panel}) and AS (\textit{right panel}) models: $\Omega_{\gamma}(a)$ (solid), $\Omega_{\rm c}(a)$ (short--dashed), $\Omega_{\rm b}(a)$ (long--dashed) and $\Omega_{\phi}(a)$ (dot--dashed).  The black (red) lines are for $\beta=0 (0.1)$.
Note that $z_{\text{eq}}$ occurs earlier (later) in the coupled 2EXP(AS) cosmology as a result of its higher (lower) CDM density at early times.}
	\label{fig:2EXPOmegas}
\end{figure*}

\noindent It is possible that potentials of this type could arise in string theory as a result of Kaluza--Klein type compactification~\cite{barreiro}. Throughout this study, we set the parameters in this potential to $\lambda_{1}=20.5$, $\lambda_{2}=0.5$ to provide $w_{\phi}\sim-0.95$ and $\sigma_{8}\sim0.8\,h^{-1}\text{Mpc}$ today for $\beta=0$. In the left panel of Fig.~\ref{fig:2EXPOmegas} we show the evolution of energy densities, $\Omega_{i}(a)$ as a function of scale factor $a$, for an uncoupled ($\beta=0$) and a coupled ($\beta=0.1$) model. The most apparent feature here is that the epoch of matter--radiation equality $z_{\text{eq}}$ occurs earlier in the coupled model. This is due to the continuous flow of energy from the CDM to the quintessence field, i.e., $\beta\dot{\phi}<0$ and so if we evolve each model to the same cosmological parameters today, the energy density of CDM for the coupled 2EXP model will be higher at earlier times relative to the uncoupled one. In the right panel of Fig.~\ref{fig:EoS} we show the redshift evolution of the Hubble parameter $H(z)$ for a coupled and uncoupled 2EXP model, which we compare to $\Lambda$CDM.

For $\beta<0$, a shallow minimum is introduced to the effective potential, allowing $\dot{\phi}$ to change sign during its evolution. This minimum is shallow enough that the field only undergoes a single oscillation, inducing  damped oscillations in the equation of state $w_{\phi}$ as can be seen from the left panel of Fig.~\ref{fig:EoS}. Hence, the direction of energy exchange between the two coupled dark components is reversed during cosmic evolution. When normalised to common cosmological parameters, the net result is that for $\beta<0$ the CDM density scales less rapidly than the uncoupled case and the energy density of CDM will be lower at earlier times. As a consequence of the minimum in the effective potential the 2EXP model with $\beta<0$ also demonstrates some early dark energy behaviour. \\

\textit{Albrecht Skordis (AS)} -- Albrecht and Skordis~\cite{albrecht} proposed a phenomenologically motivated polynomial coefficient in their scaling potential,

\begin{equation}
	V(\phi)=V_{0}[ A + (\phi - B)^{2} ]e^{-\lambda\phi} \,,
	\label{ASpotential}	
\end{equation}

\noindent which introduces a local minimum to the potential, in which the quintessence field can get trapped. This model is able to satisfy a number of cosmological constraints and provide late time acceleration, whilst keeping all the parameters in the potential $\sim \mathcal{O}(\text{M}_{\text{pl}})$. We set $\lambda=8.0$, $A=0.01$ and $B=33.960$ throughout this study, values chosen to provide $\Omega_{\phi}\sim0.7$ today when $\beta=0$ and $V_{0}\sim \mathcal{O}(\text{M}_{\text{pl}})$. With these parameter values, the field initially has a large value of $V(\phi)$ and rapidly approaches an attractor solution where $w_{\phi}$ mimics the equation of state of the dominant background energy density. Soon after the field enters the matter era, where $w_{\phi}$ should go to zero, the field reaches the minimum of the potential and undergoes small damped oscillations about that minimum (see Fig.~\ref{fig:EoS}). Due to damping from the Hubble expansion, these oscillations eventually stop and the field settles in the minimum, behaving as a cosmological constant with $w_{\phi}=-1$. These late time oscillations of the quintessence field are further damped when the field is coupled to cold dark matter with $\beta>0$; the effective potential will have a much sharper minimum, in which case the field remains in the attractor regime for a longer period, before suddenly getting trapped in the minimum without oscillating. For progressively stronger couplings, the oscillations in the field become more damped. For all values of $\beta$ considered in this study, the amplitude of the oscillations in the field are small enough that they have a negligible effect on the evolution of the CDM density. For the AS model, $H(z)$ is identical to $\Lambda$CDM over the redshift range of interest, for all coupling strengths considered in this study.

In the right panel of Fig.~\ref{fig:2EXPOmegas} we show the evolution of energy densities for an uncoupled ($\beta=0$) and a coupled ($\beta=0.1$) model. Since $\beta\dot{\phi}>0$ (the direction of energy transfer is from the quintessence scalar field to CDM) the energy density of CDM will be lower at earlier times relative to the uncoupled one if we evolve each model to the same cosmological parameters today. Hence, the epoch of matter--radiation equality $z_{\text{eq}}$ occurs later in the coupled model relative to the uncoupled one. Notice also the non--negligible contribution from dark energy at early times. With the above parameter values $A$, $B$ and $\lambda$ in the potential, it is not possible to realise a cosmology with $\Omega_{\phi}\sim0.7$ today for $\beta \lesssim-0.05$ and so we only consider $\beta>0$ for the AS models.  \\


\textit{Supergravity (SUGRA)} -- As the vacuum expectation value of the quintessence field is of the order the Planck mass, a number of authors~\cite{brax, copelandSUGRA} have argued that realistic quintessence models should receive supergravity (SUGRA) corrections. 
Such a model was proposed in~\cite{brax}, with a potential of the form

\begin{equation}
	V(\phi)=\frac{V_{0}^{4+\lambda}}{\phi^{\lambda}}e^{\phi^{2}/2} \,,
	\label{SUGRApotential}	
\end{equation}

\noindent where the evolution of $\phi$ is insensitive to a wide range of initial conditions. Throughout this work we use $\lambda=11$ to drive $w_{\phi}$ close to $-1$ today for $\beta=0$ resulting in $V_{0}\sim10^{-8}\,M_{{\rm pl}}$. Initially the value of the field is small in comparison to the Planck mass, and so the exponential factor only plays a role at low redshifts, where its presence reinforces the domination of the potential energy over the kinetic one, which pushes $w_{\phi}$ toward $-1$ today. This is demonstrated in Fig.~\ref{fig:EoS} where we show the evolution of $w_{\phi}$ for an uncoupled and coupled ($\beta=0.1$) model.

\begin{figure}[t]
	\includegraphics[width=8.5cm]{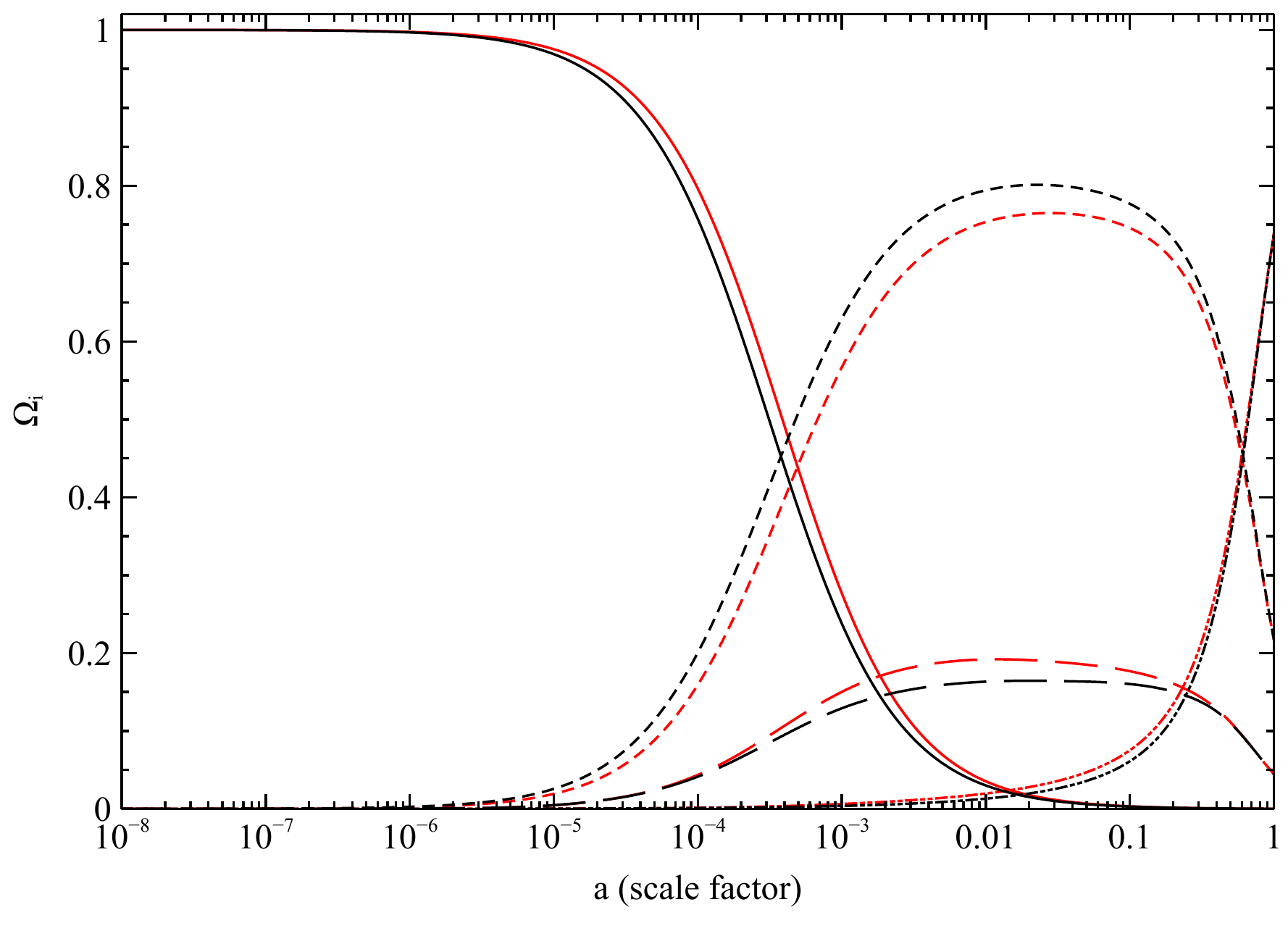}
	\caption{The cosmological evolution of the energy density parameters $\Omega_{i}$ for the SUGRA potential. Line types as in Fig.~\ref{fig:2EXPOmegas}.
Notice that $z_{{\rm eq}}$ occurs later in the coupled cosmology as a result of lower CDM density at early times.}
\label{fig:SUGRAOmegas}
\end{figure}

\begin{figure*}[t]
	\begin{tabular}{cc}
		\includegraphics[width=8.6cm]{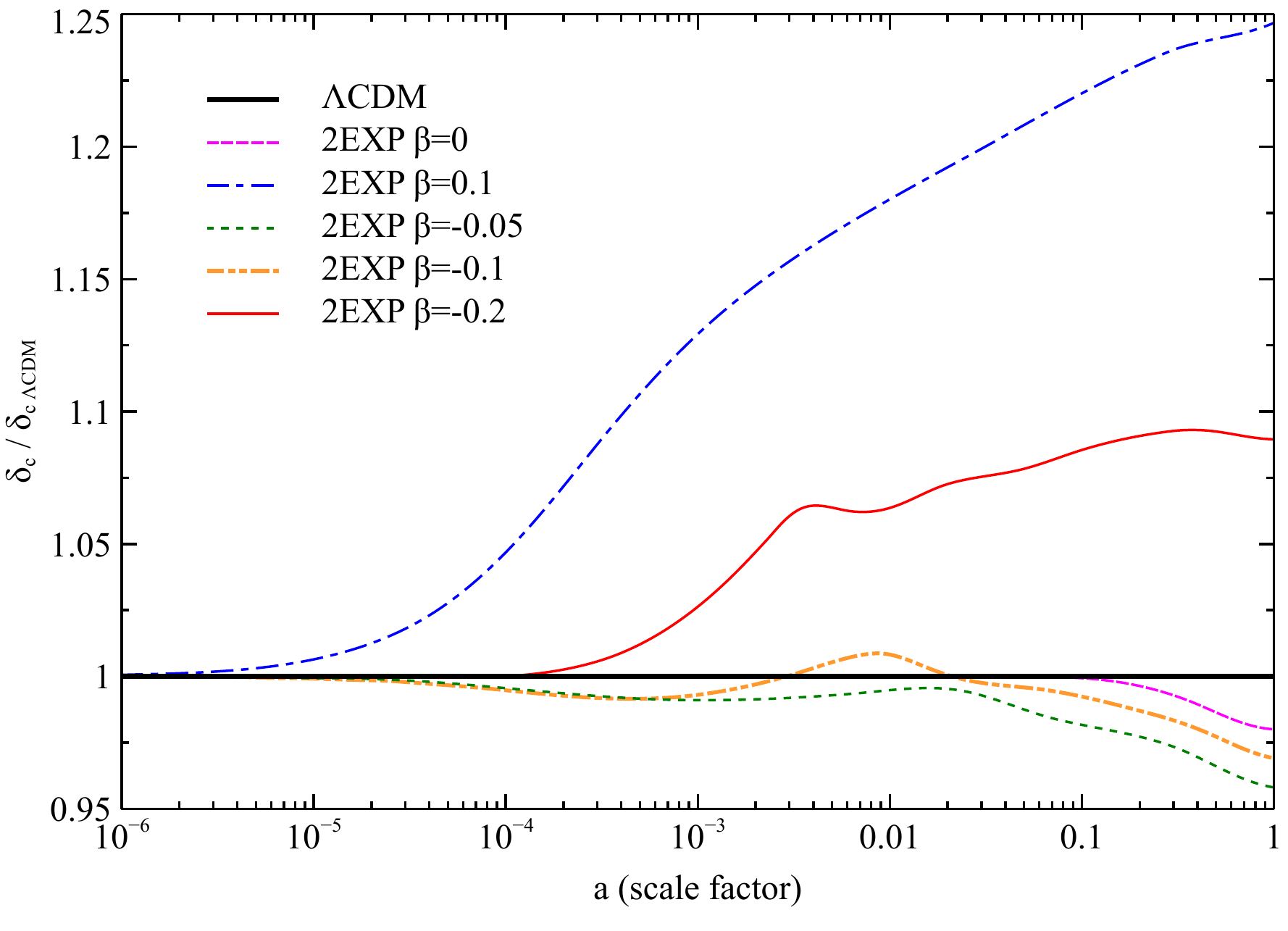} &
		\includegraphics[width=8.6cm]{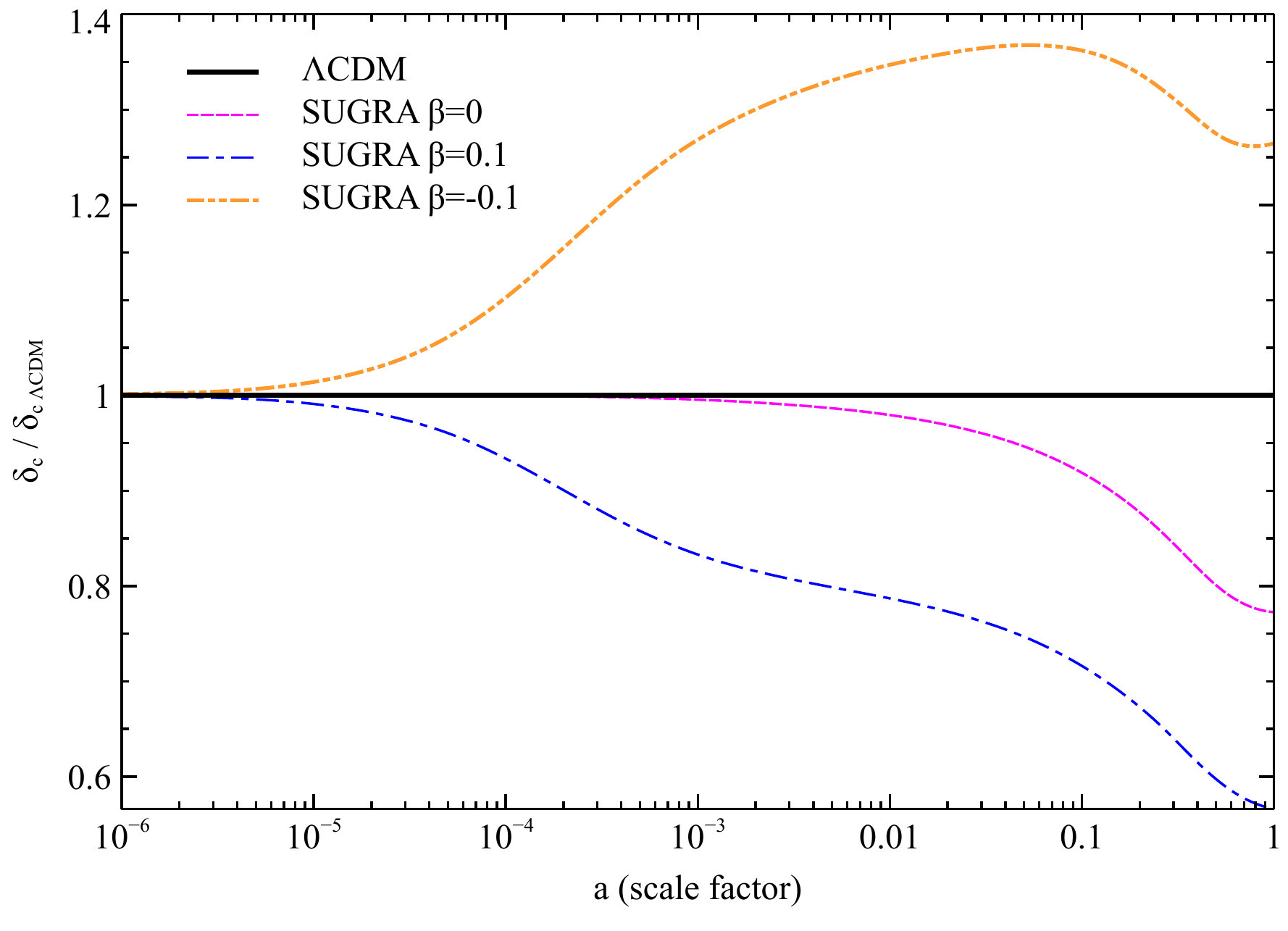}
	\end{tabular}
	\caption{Evolution of the ratio of the CDM density contrast to $\Lambda$CDM, $\delta_{{\rm c}}/\delta_{{\rm c}\,\Lambda{{\rm CDM}}}$, for the coupled 2EXP (left panel) and SUGRA (right panel) models. These figures were obtained by numerically integrating Eqs.~(\ref{bgrowthrateCQ}) and (\ref{CDMgrowthrateCQ}). Notice that the scales are different.}
	\label{fig:CDMgrowthRate}
\end{figure*}

In Fig.~\ref{fig:SUGRAOmegas} we depict the evolution of energy densities for an uncoupled ($\beta=0$) and a coupled ($\beta=0.1$) SUGRA model. Here, $z_{{\rm eq}}$ occurs later in the coupled model relative to the uncoupled one, since the evolution of $\beta\dot{\phi}>0$, i.e., the direction of energy transfer is from the quintessence scalar field to CDM. For $\beta<0$ the situation is reversed; $\beta\dot{\phi}<0$ and the energy density of CDM will be higher at earlier times relative to the uncoupled with $z_{{\rm eq}}$ occurring earlier.  The SUGRA models also have a non--negligible DE contribution at early times. In the right panel of Fig.~\ref{fig:EoS} we show the redshift evolution of the Hubble parameter $H(z)$ for a coupled and uncoupled SUGRA model.\\

In summary, the direction of energy transfer, that is, the sign of the source term $Q_{(c)0}=-\beta\rho_{\rm c}\dot{\phi}$, is dictated by the sign of $\beta$ and shape of the effective potential in which the quintessence field evolves. In a coupled model where $\beta\dot{\phi}<0$ ($>0$), $\rho_{c}$ scales less (more) rapidly than $\beta=0$ and therefore if we normalise to the same cosmological parameters today, the energy density of CDM will be higher (lower) at earlier times. This normalisation to common cosmological parameters at redshift $z=0$ allows a direct comparison between different models.


\subsection{Linear perturbations}\label{perturbations}

In this section we discuss perturbation evolution in the coupled 2EXP, SUGRA and AS models, comparing them to uncoupled models and to $\Lambda$CDM. \\

In the matter dominated epoch and on sub--horizon scales, whilst perturbations to the uncoupled baryonic component, $\delta_{\rm b}\equiv\delta\rho_{\rm b}/\rho_{\rm b}$, obey (cf. Ref.~\cite{amendolab})

\begin{equation}
		\delta''_{\rm b}=-\mathcal{H}\delta'_{\rm b}+\frac{3}{2}\mathcal{H}^{2}\left[\Omega_{\rm c}\delta_{\rm c}+\Omega_{\rm b}\delta_{\rm b}\right] \,,
	\label{bgrowthrateCQ}	
\end{equation}

\noindent perturbations to the coupled CDM component, $\delta_{\rm c}\equiv\delta\rho_{\rm c}/\rho_{\rm c}$, follow

\begin{equation}
		\delta''_{\rm c}=-(\mathcal{H}+\beta\phi^{\prime})\delta'_{\rm c}+\frac{3}{2}\mathcal{H}^{2}\left[\Omega_{\rm c}\delta_{\rm c}\frac{G_{{\rm  eff}}}{G_{{\rm N}}} + \Omega_{\rm b}\delta_{\rm b}\right] \,,
	\label{CDMgrowthrateCQ}	
\end{equation}

\noindent with

\begin{equation}
		G_{{\rm  eff}}=G_{{\rm N}}(1+2\beta^{2}) \,,
	\label{Geffective}	
\end{equation}

\noindent where $G_{{\rm N}}$ is Newton's constant. Here, the prime denotes differentiation with respect to conformal time and $\mathcal{H}=a^{\prime}/a$ is the conformal Hubble function. In the expressions above we have assumed that the scalar field is massless, which is a very good approximation for the models we consider here, since the length scales we are interested in are much smaller than the interaction range of the quintessence field.

For $\beta\neq0$, the growth of cold dark matter perturbations are affected in a number of ways~\cite{baldi0,liBarrow,baldic}. Firstly, the background expansion rate $H(z)$ is modified, which will lead to faster or slower clustering of matter particles. Secondly the friction term of Eq.~(\ref{CDMgrowthrateCQ}) is modified by the extra contribution, $\beta\phi^{\prime}$, implying that the rate of perturbation growth will then depend both on the coupling $\beta$ and the background evolution of the scalar field. Secondly, the CDM particles feel an additional `fifth--force' (an effective Newton's constant $G_{{\rm eff}}$) which is mediated by the scalar field. Finally, the coupling effectively rescales the CDM particle mass which the uncoupled baryonic component will feel through a changing gravitational potential generated by the CDM. 

In Fig.~\ref{fig:CDMgrowthRate} we show the evolution of the ratio of the CDM density contrast in the coupled 2EXP and SUGRA models to $\Lambda$CDM, $\delta_{\rm c}/\delta_{{\rm c}\,\Lambda{\rm CDM}}$. These figures were obtained by numerically integrating Eqs.~(\ref{bgrowthrateCQ}) and (\ref{CDMgrowthrateCQ}). For each model, the same initial conditions $\dot{\delta}^{i}_{{\rm c,b}}$ and $\delta^{i}_{{\rm c,b}}$ were used and the integration was started at a high redshift $(1+z_{i}=10^{9})$ to minimise the effect of the decaying mode on the evolution of the perturbations.

For the 2EXP model with $\beta>0$, the $\beta\phi^{\prime}$ term contributes negatively to the total friction term of Eq.~(\ref{CDMgrowthrateCQ}) which, along with the additional fifth force and higher CDM density at early times, leads to an enhancement of growth with respect to $\Lambda$CDM. For $\beta=0$, the presence of the scalar field leads to a suppression of perturbation growth. We also find a suppression of growth in the uncoupled SUGRA and AS models. This is a well understood feature of uncoupled quintessence~\cite{joyce,caldwell,jennings}; just as fluctuation growth is suppressed at late times with the onset of dark energy, the growth of linear perturbation modes are slowed at early times due to a non--negligible fraction of quintessence scalar field energy density.

Despite the presence of the fifth force, CDM perturbation growth in the SUGRA and AS models is suppressed further for $\beta>0$ with respect to $\Lambda$CDM due to the lower CDM density at early times and since $\beta\phi^{\prime}$ contributes positively to the total friction term. For $\beta<0$, growth in the SUGRA model is enhanced compared to the uncoupled case and exceeds $\Lambda$CDM for progressively more negative couplings, due to the combined effect of the fifth force and since the $\beta\phi^{\prime}$ term is anti--frictional. This behaviour is also true of the AS model. For $\beta<0$ in the 2EXP models, the growth of CDM perturbations has a more complicated dependence on the magnitude of $\beta$, due to the reversed direction of energy exchange between the CDM and quintessence field, induced by the shallow minimum in the effective potential. For weak, negative couplings $\beta\sim-0.05$, the net effect is a suppression of growth relative to $\Lambda$CDM and the uncoupled 2EXP model. When the coupling strength is increased to $\beta\sim-0.1$, $\beta\dot{\phi}<0$ as the field rolls toward the minimum of the effective potential. This contributes negatively to the total friction term of Eq.~(\ref{CDMgrowthrateCQ}), which combined with the fifth force acts to slightly enhance perturbation growth at early times ($z\sim100$) with respect to $\Lambda$CDM and the uncoupled 2EXP model. Once the field turns over, $\beta\dot{\phi}>0$ in competition with the fifth force, with the net result that perturbation growth is suppressed at late times ($z\sim20$) relative to $\Lambda$CDM and the uncoupled 2EXP model, but enhanced relative to $\beta=-0.05$.  For even stronger coupling strengths $\beta\sim-0.2$, the effect of the small frictional term at early times and the fifth force result in a enhancement of growth relative to $\Lambda$CDM. In practice, the impact of these individual effects on cosmic structure formation is not always obvious; one mechanism may act to enhance structure formation, whilst another may suppress it (cf. Ref.~\cite{baldic}).

\begin{figure*}[t]
	\begin{tabular}{ccc}
		\includegraphics[width=5.9cm]{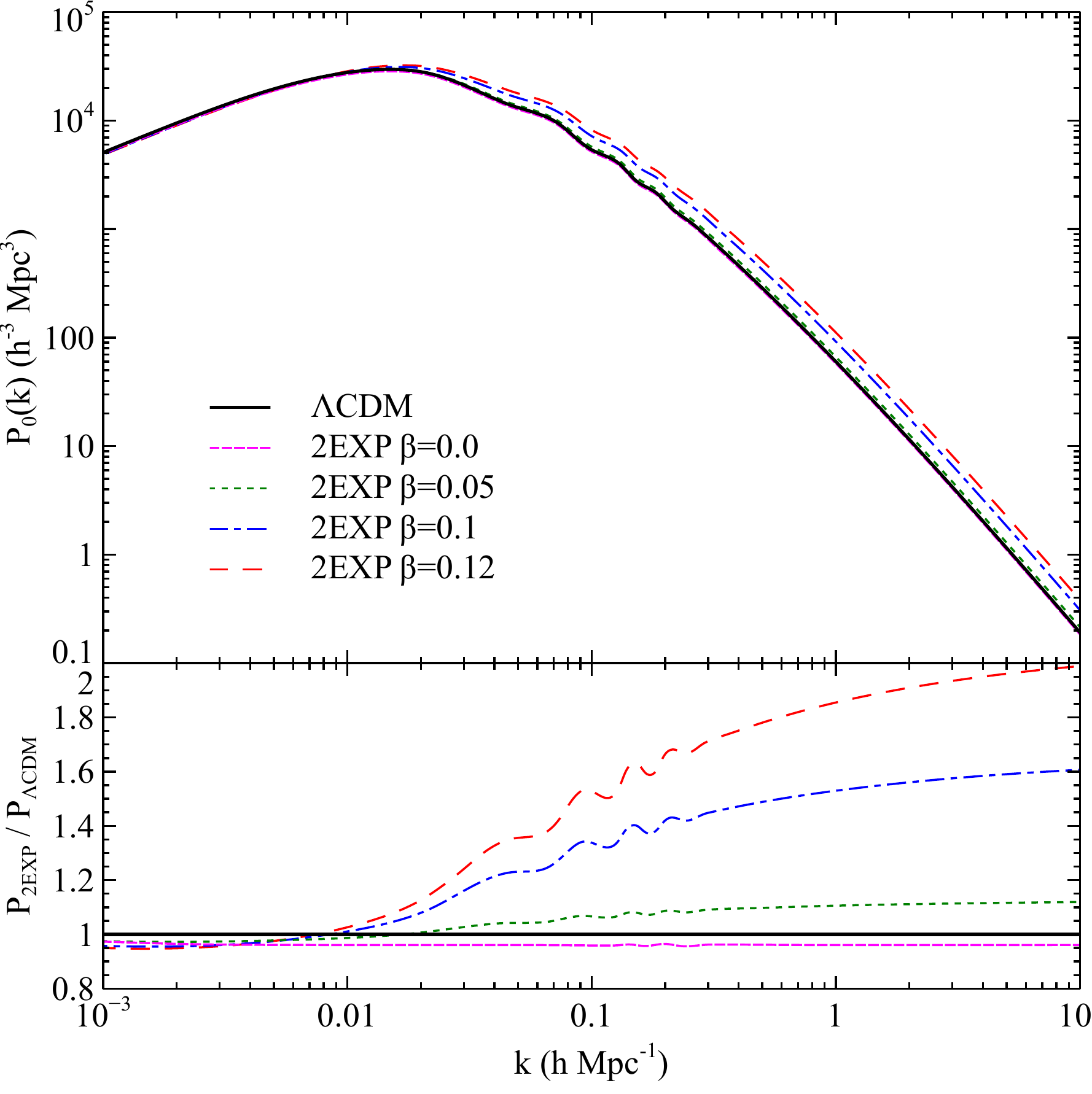} &	
		\includegraphics[width=5.9cm]{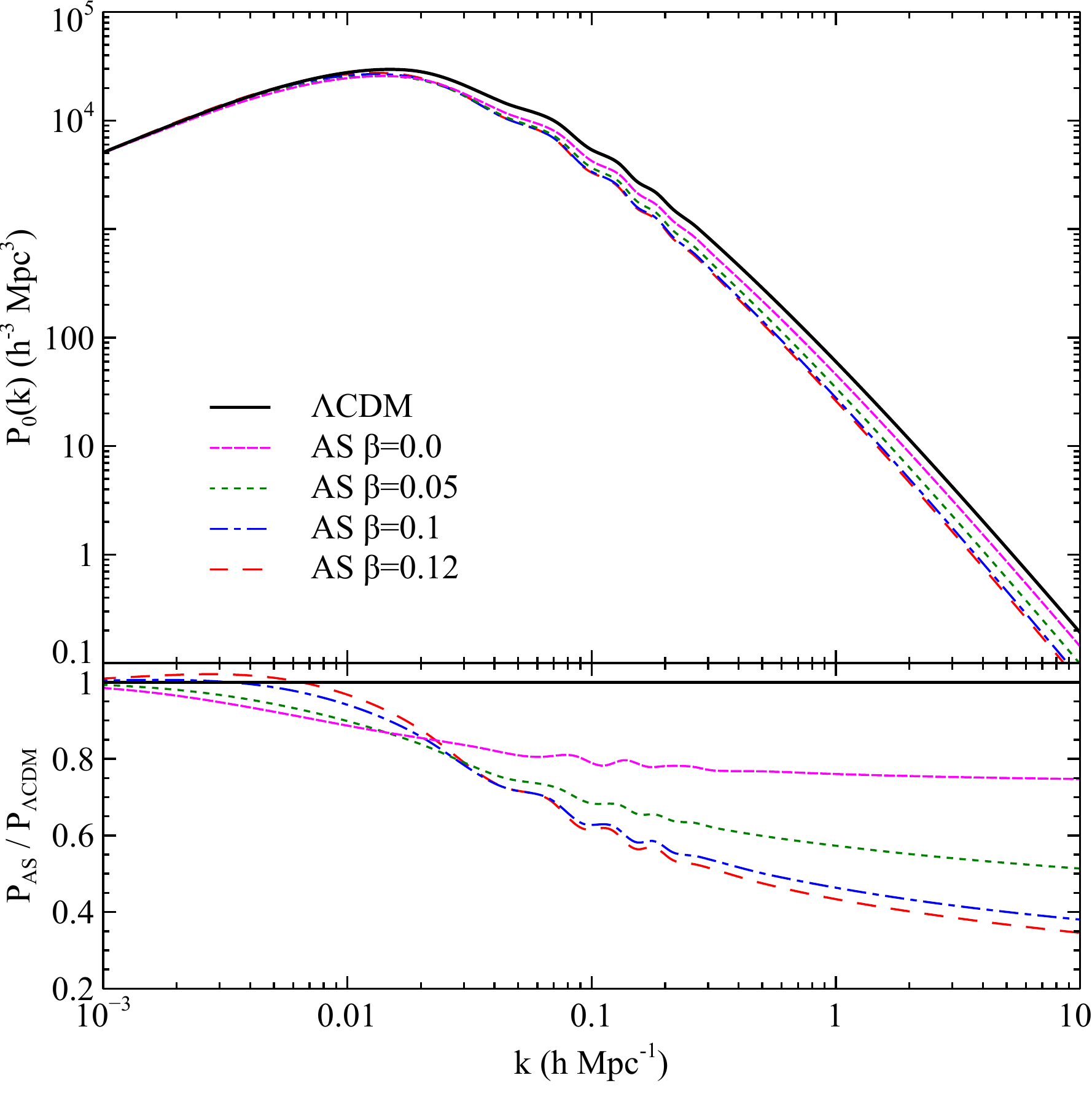} &
		\includegraphics[width=5.9cm]{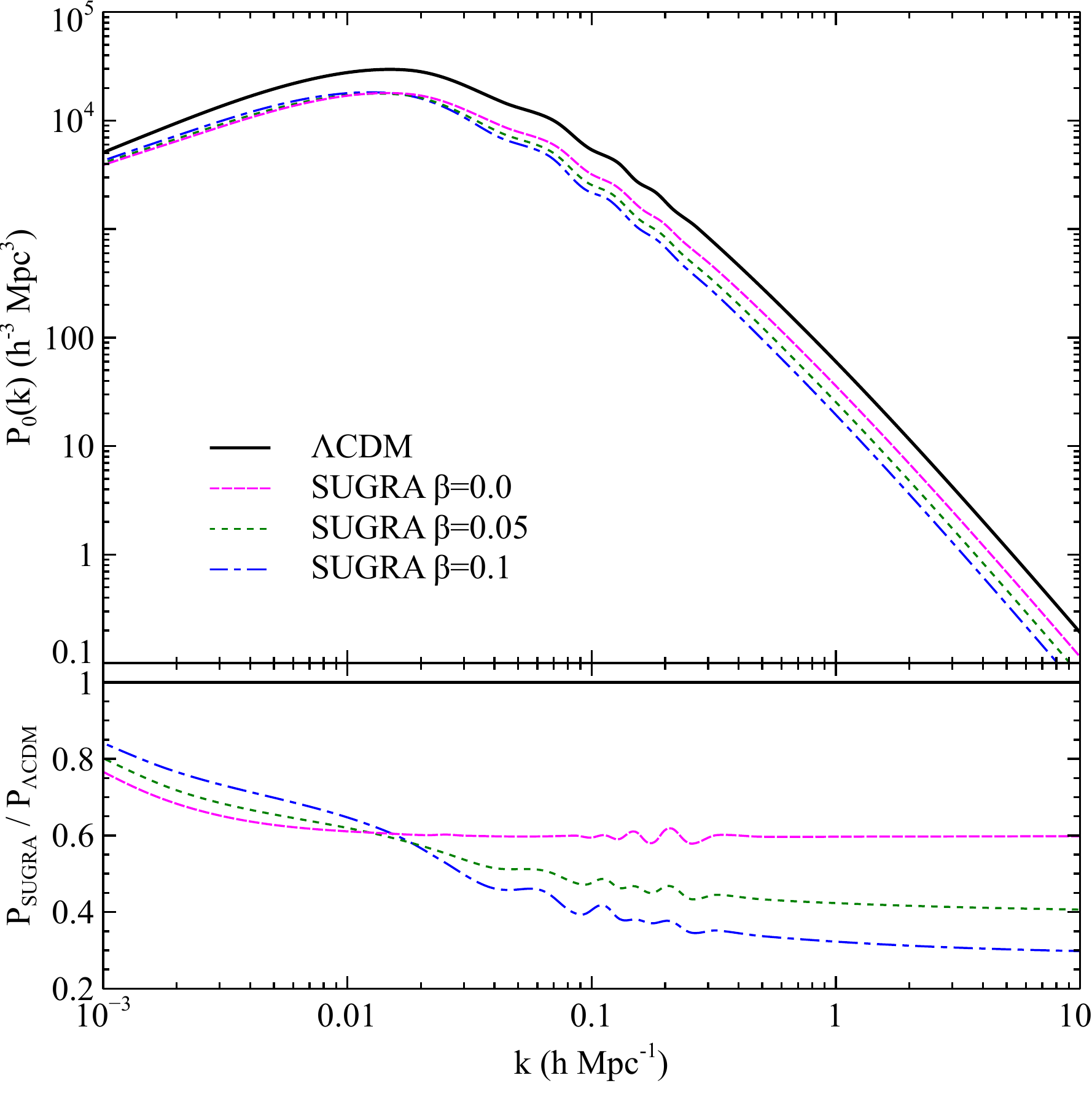}
	\end{tabular}
	\caption{The linear (total) matter power spectrum at $z=0$, $P_{0}(k)$, generated using the \texttt{CAMB} computer software for the models under consideration. We show various coupling strengths and compare to the linear power spectrum of $\Lambda$CDM in the lower panels. \textit{Left panel}: 2EXP model. \textit{Middle panel}: AS model. \textit{Right panel}: SUGRA model. Note that all spectra have been normalised to CMB fluctuations, which are on larger scales than included in the figure.}
	\label{fig:matterPowerSpectrum}
\end{figure*}

It is clear that changes in the background expansion history of the universe and modifications to the growth rate of the matter perturbations will affect the matter power spectrum. In Fig.~\ref{fig:matterPowerSpectrum} we plot the linear (total) matter power spectrum at $z=0$, $P_{0}(k)$, generated using \texttt{CAMB}. All the spectra have been normalised to CMB fluctuations. The left, middle and right panels show the power spectra for the 2EXP, AS and SUGRA models respectively for a variety of positive coupling strengths $\beta$. For the 2EXP model, when the coupling strength is increased, small scale power is enhanced and a slight suppression of power on large scales is observed. This is because the epoch of matter--radiation equivalence ($z_{{\rm eq}}$) occurs earlier when $\beta>0$ and so only the very small scale perturbations on scales $k>k_{{\rm eq}}$ have time to enter the horizon and grow during radiation domination. Hence, the turnover in the power spectrum (which corresponds to the scale that entered the horizon when the universe became matter dominated) occurs on smaller scales and since the growth of perturbations for $\beta>0$ are enhanced, small scale power is increased. For fluctuations on scales $k<k_{{\rm eq}}$ during the matter dominated epoch, the enhancement or suppression takes place after the mode enters the horizon. Consequently, the amplitude of the linear power spectrum on $8\,h^{-1} {\rm Mpc}^{-1}$ scales ($\sigma_{8}$) is higher compared to $\Lambda$CDM, whilst the uncoupled 2EXP model has a lower $\sigma_{8}$ compared to $\Lambda$CDM when normalised to CMB fluctuations. Similar arguments may be applied to explain the suppression/enhancement of power seen on different scales for the AS and SUGRA cosmologies (middle and right panels of Fig.~\ref{fig:matterPowerSpectrum}). \\


\section{Abundance of CDM halos}\label{massfunction}

While N--body simulations are a powerful tool for studying cosmic structure formation, they can be very time consuming. Therefore semi--analytic methods are a useful alternative, in particular for exploring a range of non--standard cosmological models involving a wide parameter space. The semi--analytical expressions used below have been shown to be in reasonable agreement with N-body simulations at calculating the abundance of dark matter halos over a wide range of masses, however we note that the expressions for the mass functions used below have been tested only against N--body simulations of $\Lambda$CDM cosmologies and not for coupled quintessence models. Fortunately, N--body simulations presented in \cite{baldi0} showed evidence that the Jenkins et al. mass function \cite{jenkins} represents a good fit to the mass functions found numerically even in coupled quintessence models. 

\subsection{Mass functions}

Observations indicate that individual galaxies and clusters of galaxies are embedded in extended halos of dark matter. Press and Schechter~\cite{press} were the first to provide a theoretical description for the abundance of these halos as a function of their mass using the assumption that the fraction of the volume of the universe that has collapsed into objects of mass $M$ at a redshift $z$ follows a Gaussian distribution:

\begin{equation}
	\mathcal{F}_{\text{col}}(M,z)=\frac{2}{\sqrt{2\pi}\sigma(R,z)}\int^{\infty}_{\delta_{\star}(z)}{\rm d}\delta \, e^{-\delta^{2}/2\sigma^{2}(R,z)} \,,
	\label{PressSchecterSTART}	
\end{equation}
\noindent where $\delta_{\star}(z)$ is the critical density which an object must exceed for collapse to occur at the redshift $z$ of interest. The density field has been smoothed on a comoving scale, $R$, and $\sigma^{2}(R,z)$ is the mass variance:

\begin{equation}
		\sigma^{2}(R,z)=\frac{D^{2}_{+}(z)}{2\pi^{2}}\int^{\infty}_{0} {\rm d}k\, k^{2}P_{0}(k)W^{2}(k;R)\,,
		\label{sigmaVariance}	
\end{equation}

\noindent where $P_{0}(k)$ is the linear matter power spectrum today and $W(k;R)$ is the Fourier transform of the window function used to smooth the density field. The redshift dependence of the power spectrum has been extracted by multiplying by the linear growth factor of the perturbations of the total matter ($\rho_{\rm m}=\rho_{\rm c}+\rho_{\rm b}$) perturbations, normalised to unity today:

\begin{equation}
		D_{+}(z)\equiv\frac{\delta_{\rm m}(z)}{\delta_{\rm m}(z=0)} \,,
	\label{growthFactor}	
\end{equation}

\noindent where $\delta_{\rm m}\equiv\delta\rho_{\rm m}/\rho_{\rm m}$. For a spherical top--hat filter of radius $r$ in real space, containing a mass

\begin{equation}
		M(r,z)=\frac{4}{3}\pi\frac{\bar{\rho}_{\rm m}(z)}{(1+z)^{3}}r^{3} \,,
		\label{RMrelation}	
\end{equation}

\noindent where $\bar{\rho}_{\rm m}(z)$ is the mean total matter density at redshift $z$, then

\begin{equation}
		W(k;R)=3\left(\frac{\text{sin}(kR)}{(kR)^{3}} - \frac{\text{cos}(kR)}{(kR)^{2}}\right) \,.
		\label{window}	
\end{equation}

\noindent To obtain the comoving number density of objects of mass between $M$ and $M+{\rm d}M$ at redshift $z$, we first differentiate Eq.~(\ref{PressSchecterSTART}) with respect to $M$ which gives the fraction of the volume in the universe that has collapsed into objects with mass between $M$ and $M+{\rm d}M$. Multiplying this by $\bar{\rho}_{\rm m}(z)(1+z)^{-3}/M$ gives the comoving number density of collapsed objects with mass between $M$ and $M+{\rm d}M$ at redshift $z$:

\begin{equation}
		\frac{{\rm d}n(M,z)}{{\rm d}M}=\frac{1}{M}\frac{\bar{\rho}_{\rm m}(z)}{(1+z)^{3}}\frac{{\rm d}\mathcal{F}_{{\rm col}}}{{\rm d}M}\,.
		\label{comovingNumberDens}	
\end{equation}

\noindent Carrying out the differentiation gives

\begin{eqnarray}
  && \frac{{\rm d}n(M,z)}{{\rm d}M} = \sqrt{\frac{2}{\pi}}\frac{1}{M}\frac{\bar{\rho}_{{\rm m}}(z)}{(1+z)^{3}}\frac{\delta_{\star}(z)}{\sigma(R,z)}\, \nonumber \\
		 && \times \, \text{exp}\left(-\frac{\delta^{2}_{\star}(z)}{2\sigma^{2}(R,z)}\right)\left[-\frac{1}{\sigma(R,z)}\frac{{\rm d}\sigma(R,z)}{{\rm d}M}\right] \,,
		\label{diffMassFunction}	
\end{eqnarray}

\noindent and is referred to as the differential mass function. Writing the term in the square brackets as the logarithmic derivative ${\rm d}\,\ln{ \sigma^{-1}}/{\rm d}M$, it is common practice to re--write Eq.~(\ref{diffMassFunction}) as 

\begin{equation}
  f(\sigma,z)\equiv M\frac{(1+z)^{3}}{\bar{\rho}_{m}(z)}\frac{{\rm d} n(M,z)}{{\rm d} \ln{\sigma^{-1}}} \,,
		\label{massFunctionDefn}	
\end{equation}

\noindent which defines the mass function.  Hence, (dropping the arguments of $\delta_{\star}(z)$ and $\sigma(R,z)$ for compactness) the PS mass function is

\begin{equation}
  f(\sigma;\text{PS})=\sqrt{\frac{2}{\pi}}\frac{\delta_{\star}}{\sigma}\,\exp{\left[-\frac{\delta^{2}_{\star}}{2\sigma^{2}}\right]} \,,
		\label{PressShectMF}	
\end{equation}

\noindent and so has an implicit dependence on redshift, through the parameter $\delta_{\star}(z)$. The PS mass function is normalised so that

\begin{equation}
  \int^{\infty}_{-\infty}f(\sigma;\text{PS}) \, {\rm d} \ln{\sigma^{-1}}=1 \,,
		\label{PressShecNormalisation}	
\end{equation}

\noindent which implies that all of the matter is within haloes of some mass. It has been found to over--predict(under--predict) the number of low(high) mass haloes at the current epoch~\cite{jenkins,marassi}. Sheth and Tormen~\cite{sheth} (hereafter ST) 
found that this discrepancy can be substantially reduced by considering ellipsoidal, rather than
spherical collapse. Their mass function contains three empirically determined parameters $A$, $a$ and $p$:

\begin{equation}
  f(\sigma;\text{ST})=A\sqrt{\frac{2a}{\pi}}\left[1+\left(\frac{\sigma^{2}}{a\delta^{2}_{\star}}\right)^{p}\right]   	
          \frac{\delta_{\star}}{\sigma}\,\text{exp}\left[-\frac{\delta^{2}_{\star}}{2\sigma^{2}}\right] \,.
		\label{ShethTormenMF}	
\end{equation}

\noindent We use $A=0.322$, $a=0.707$ and $p=0.3$~\cite{sheth}, for all the quintessence models considered. The functional form of $f$ in Eq.~(\ref{massFunctionDefn}) has also been directly fitted to the mass function obtained from numerical N--body simulations as a function of $\sigma$ only. For example, Jenkins et al.~\cite{jenkins} found 

\begin{equation}
  f(\sigma;\text{Jenkins})=0.315\,\exp{(-|\ln{\sigma^{-1}}+0.61|^{3.8})} \,,
		\label{JenkinsMF}	
\end{equation}

\noindent valid over the range $-1.2\leq \ln{\sigma^{-1}} \leq1.05$. Note the range over which the fit function is valid. For standard cosmologies, this corresponds approximately to the mass range $10^{11}$ to $10^{16}\,\text{M}_{\odot}$ at $z=0$. We find that for some combinations of quintessence potential and coupling, the mass/redshift range over which this fit function is valid is dramatically reduced. Since Eq.~(\ref{JenkinsMF}) depends on the variance $\sigma$ only, it is manifestly independent of cosmology--specific parameters such as redshift and critical density for collapse. This is what is commonly understood as the `\textit{universality of the mass function}', (see Refs.~\cite{alimi} and \cite{pierStefano} for recent discussions of this topic). Mass function universality suggests that semi--analytic forms such as ST and fit functions like Jenkins should remain valid in coupled quintessence cosmologies. Indeed, this has been verified by a number of authors~\cite{baldi0,baldib} by using N--body simulations. Another mass function, similar in form to ST has been introduced
by Tinker et al.~\cite{tinker}. It has been calibrated over a wide mass and redshift range using N--body simulations, but is however, still only valid over a limited range in $\ln{\sigma^{-1}}$. Hence, for our non--standard CQ models, the Tinker mass function offers no real advantages over ST, since to probe high ($z>1$) redshifts would require extrapolating the Tinker fit--function beyond its range of validity. Recently, a path--integral method that allows for an analytical computation of the mass function for more generic filters than the one described above, has been used to provide a robust theoretical estimation of the halo mass function~\cite{pierStefano}. Based on the Path Integral approach applied recently to the Excursion Set formalism of Ref.~\cite{Maggiore2009rv}, these theoretical predictions show a remarkable agreement with the numerical results of Tinker et al. with deviations of no greater than 5\% over the range of masses probed by the simulations. This suggests that the Excursion Set formalism, combined with a realistic model of the dark matter halo collapse conditions can provide a universal formulation of the halo mass function. Although, not the thrust of this piece of work, it would certainly be worth following up this possibility for the case of coupled quintessence.\\

A cumulative number count $\mathcal{N}(>M,z,\Delta z)$, above a given mass threshold $M=M_{\rm th}$, in a redshift slice ${\rm d}z$ is obtained from ${\rm d}n(M,z)/{\rm d}M$:

\begin{equation}
   	\mathcal{N}(>M,z,\Delta z) = \Delta\Omega \int^{z+\Delta z}_{z} {\rm d}z \frac{{\rm d}V}{{\rm d}z} n(>M,z) \,, 
	\label{numberCount}	
\end{equation}

\begin{equation}
	n(>M,z)=\int^{\infty}_{M_{\rm th}} {\rm d}M\frac{{\rm d}n(M,z)}{{\rm d}M} \,,
	\label{numberDensity}	
\end{equation}

\noindent where ${\rm d}V/{\rm d}z$ is the comoving volume element, which in a spatially flat universe takes the form

\begin{equation}
   	\frac{{\rm d}V}{{\rm d}z}=\Delta\Omega r^{2}(z)\frac{{\rm d}r}{{\rm d}z} \,.
	\label{volumeElement}	
\end{equation}

\noindent $\Delta\Omega$ is the solid angle and $r(z)$ denotes the comoving radial distance out to redshift $z$: 

\begin{equation}
   	r(z)=\frac{c}{H_{0}}\int^{z}_{0} \frac{{\rm d}x}{E(x)} \,,
	\label{comovAngDiaDist}	
\end{equation}

\noindent where $E(z)=H(z)/H_{0}$. This geometrical factor ${\rm d}V/{\rm d}z$ is required since the redshift evolution of a physical `volume' in space is model dependent, depending (among other things) on the coupling strength and form of the potential. For numerical computation, the upper limit of integration in Eq.~(\ref{numberDensity}) is replaced by some finite mass value $M_{{\rm max}}$.

The final ingredient required is a method for calculating the critical density contrast for collapse, $\delta_{\star}(z)$. In a matter dominated universe, this parameter is a constant $\delta_{\star}=1.686$, whilst for $\Lambda$CDM it becomes a weak function of cosmology, asymptoting to $\delta_{\star}=1.686$ at high redshift~\cite{lahav,eke}. In cosmologies with an early dark energy component however, it has been established (see Refs.~\cite{bartelmann,francis} for example) that the value of $\delta_{\star}$ can have a small, but non--negligible effect in the high mass/redshift end of the mass function, where the cluster abundance is smallest. As discussed in e.g. Refs.~\cite{bruck} and~\cite{francis}, if DE were not to remain homogeneous over the relevant length scales, or if, as is the interest here, DE is coupled to dark matter, then the DE density within a perturbation may evolve separately from the background universe. We should then expect significant deviations from $\delta_{\star}=1.686$. Determining $\delta_{\star}(z)$ in CQ cosmologies has recently been the subject of some debate~\cite{wintergerst}. In the next section, we present the spherical collapse model applied to CQ and the model proposed in Ref.~\cite{wintergerst}. Both methods are based on assumptions: In Ref.~\cite{wintergerst} for example, one neglects the contribution of the scalar field perturbation in the Poisson equation, whilst in the spherical collapse model one assumes that there is no loss in scalar field energy density from inside the collapsing region. To get a feeling for how the lack of understanding in modelling the non--linear behaviour of the scalar field manifests itself in the halo mass function, we subsequently compare these two methods for determining $\delta_{\star}(z)$.


\subsection{Critical density for collapse, $\delta_{\star}(z)$}

In this section, we denote background quantities with a bar, we assume matter domination and we neglect baryons. The evolution of a density perturbation during the non--linear phase of gravitational collapse is most simply approximated by the spherical collapse model~\cite{padmanabhanb, peacock}.  The model was first applied to the standard CDM scenario and later to $\Lambda$CDM~\cite{lahav}. Since then, the spherical collapse model has been applied to quintessence scenarios~\cite{wang,bartlemannDoran,bruck} and more recently to interacting DE--CDM models~\cite{mota}. In an alternative method, Ref.~\cite{mainini} made predictions for $\delta_{\star}(z)$ in CQ models, explicitly accounting for an uncoupled baryonic component by considering the evolution of spherical concentric shells.

\noindent All background quantities ($H,\bar{\phi},\bar{\rho}_{\rm m}$) follow the standard evolution equations for coupled quintessence (cf. Ref.~\cite{amendolab}). Defining $\delta\phi$ as the perturbation to the quintessence field, $\phi=\bar{\phi}+\delta\phi$, the non--linear evolution of the density contrast is governed by~\cite{mota}

\begin{eqnarray}
	\ddot{\delta}_{\rm m} &=& -2H\dot{\delta}_{\rm m} + \frac{1}{2}\bar{\rho}_{\rm m}\delta_{\rm m}(1+\delta_{\rm m}) 
	+ \frac{4}{3}\frac{\dot{\delta}^{2}_{\rm m}}{1+\delta_{\rm m}} \nonumber \\
																				 && + \frac{1}{2}\left[\rho_{\phi}(1+3w_{\phi})-\rho_{\bar{\phi}}
	(1+3w_{\bar{\phi}})\right](1+\delta_{\rm m}) \nonumber \\
	                                       && + \beta\left[-\frac{2}{3}\dot{\delta\phi}\dot{\delta}_{\rm m}
	+ \left(2H\dot{\delta\phi}+\ddot{\delta\phi}\right)(1+\delta_{\rm m})\right]	\nonumber \\ 
	                                       && + \beta^{2}\frac{1}{3}\dot{\delta\phi}^{2}(1+\delta_{\rm m}) \,.
	\label{modifiedSphCol}	
\end{eqnarray}

\noindent The evolution equation for the scalar field inside the overdensity reads:

\begin{equation}
	\ddot{\phi} + 3\left(H+\frac{\beta}{3}\dot{\delta\phi}-\frac{1}{3}\frac{\dot{\delta}_{\rm m}}{1+\delta_{\rm m}}\right)\dot{\phi}
				+ \frac{{\rm d}V}{{\rm d}\phi} =-\beta\bar{\rho}_{\rm m}(1+\delta_{\rm m}) \,.
	\label{sphrhophinoR}	
\end{equation}

From Eq.~(\ref{modifiedSphCol}) we see that the original non--linear equation for the density contrast is modified by the presence of the coupling terms and by a term that originates from the assumption that the quintessence field inside the overdense region evolves separately from the background.

Since we are interested in calculating the extrapolated linear overdensity of a matter perturbation collapsing at any given redshift, i.e., $\delta_{\star}(z)=\delta_{{\rm m,L}}(z=z_{\rm c})$, we solve numerically Eqs.~(\ref{modifiedSphCol}) and (\ref{sphrhophinoR}) with $\dot{\delta}^{i}_{\rm m}=0$ and search for the value of the initial overdensity $\delta^{i}_{\rm m}$ for which the collapse occurs at redshift $z_{{\rm c}}$. Formally, this is when $\delta_{{\rm m}}(z_{{\rm c}})\rightarrow\infty$, i.e., the sphere collapses toward a central singularity. Of course, in reality, dissipative processes intervene well before such a singularity is reached and the system will virialise. Having determined the initial overdensity for the redshift of interest, we use it as an initial condition to solve the linearised form of Eq.~(\ref{modifiedSphCol}) which defines the linearly extrapolated density contrast $\delta_{{\rm m,L}}$,

\begin{eqnarray}
	\ddot{\delta}_{{\rm m,L}} &=& -2H\dot{\delta}_{{\rm m,L}} + \frac{1}{2}\bar{\rho}_{{\rm m}}\delta_{{\rm m,L}}   \nonumber \\
							      && + \frac{1}{2}\big[(1+3\bar{w}_{\bar{\phi}})\rho_{\bar{\phi}}\delta_{\bar{\phi}}
	                   + 3\rho_{\bar{\phi}}\delta w_{\bar{\phi}}\big] \nonumber \\
	                  && + \beta(2H\dot{\delta\phi} + \ddot{\delta\phi}) \,,
	\label{linearOverdensity}	
\end{eqnarray}

\noindent where we have defined $\rho_{\phi}=\rho_{\bar{\phi}}(1+\delta_{\bar{\phi}})$, $\delta_{\bar{\phi}}=\delta\rho_{\bar{\phi}}/\rho_{\bar{\phi}}$. Furthermore, we have 

\begin{equation}
	\delta w_{\bar{\phi}}=(1-w_{\bar{\phi}})\left(\delta_{\bar{\phi}} - \frac{1}{V}\frac{{\rm d}V}{{\rm d}\bar{\phi}}\delta\phi\right) \,,
	\label{deltaw}	
\end{equation}

\noindent and

\begin{equation}
	\delta\rho_{\bar{\phi}} = \dot{\bar{\phi}}\dot{\delta\phi} + \frac{{\rm d}V}{{\rm d}\bar{\phi}}\delta\phi \,.
	\label{deltarhophi}	
\end{equation}

\begin{figure*}[t]
	\begin{tabular}{ccc}
		\includegraphics[width=5.9cm]{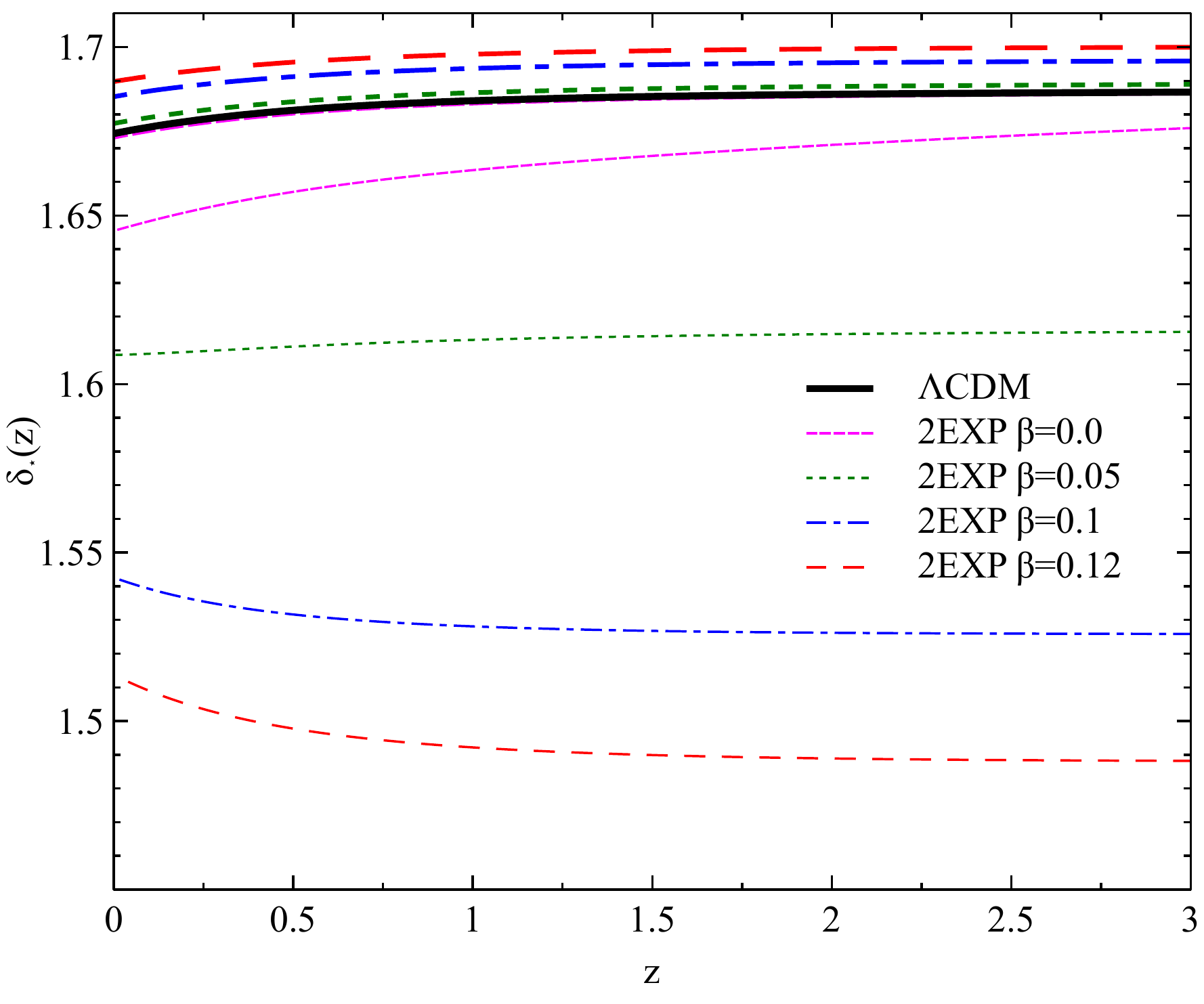} &	
		\includegraphics[width=5.9cm]{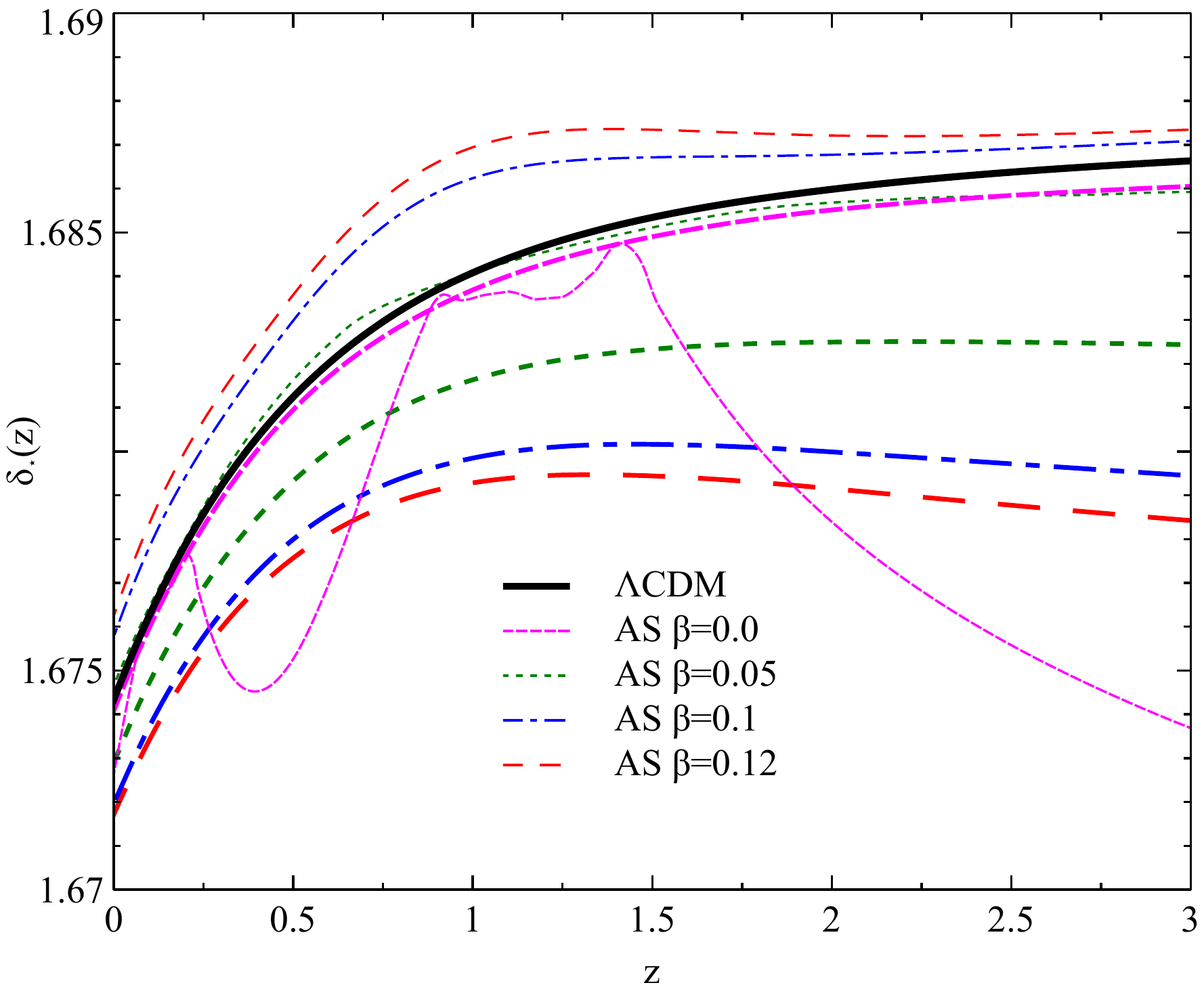} &
		\includegraphics[width=5.9cm]{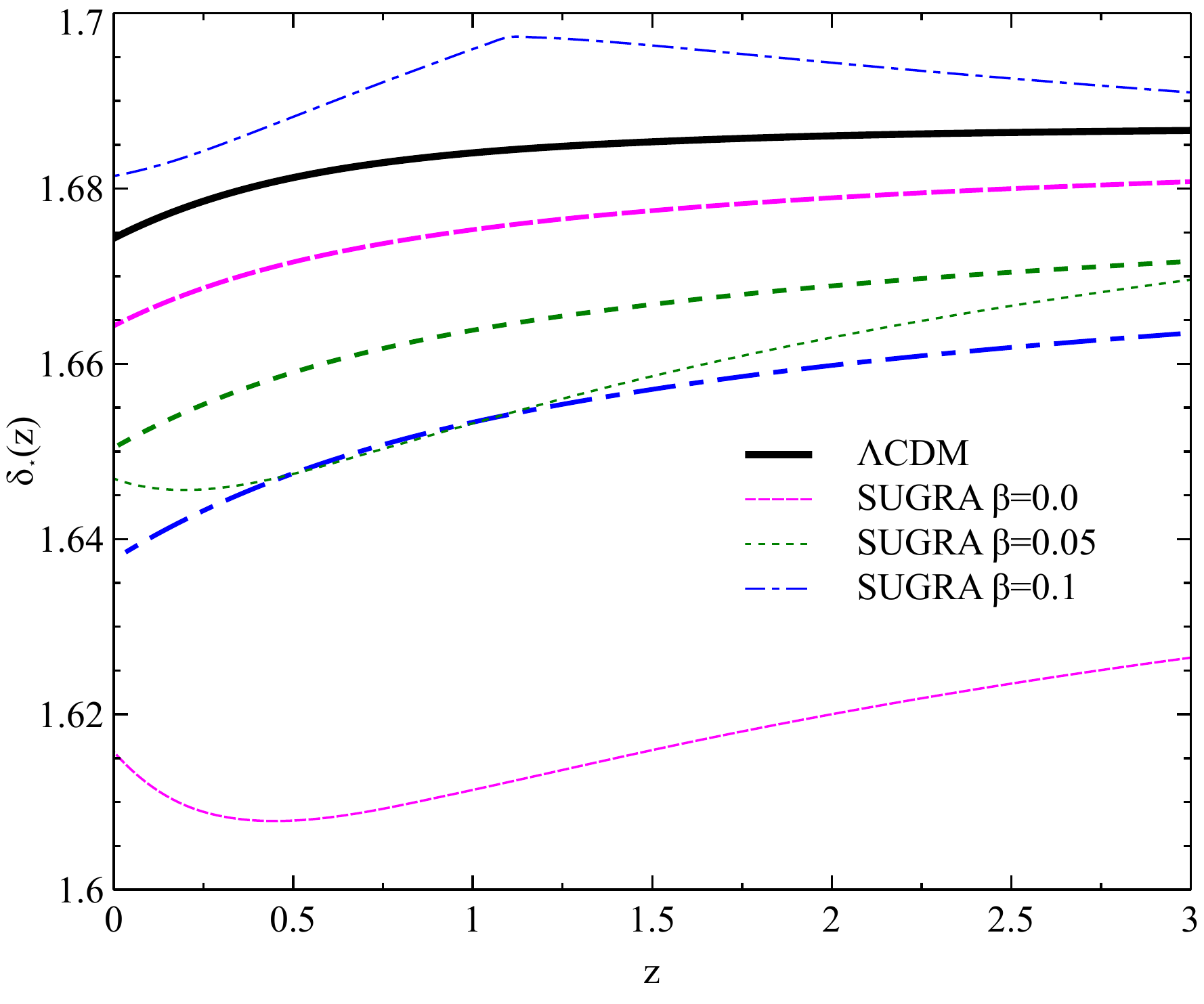}
	\end{tabular}
	\caption{The redshift evolution of the linearly extrapolated density contrast at the redshift of collapse $\delta_{\star}$. We show various coupling strengths $\beta>0$ and compare to the $\Lambda$CDM value (thick solid back line). \textit{Left panel}: 2EXP model. \textit{Middle panel}: AS model. \textit{Right panel}: SUGRA model. Notice that the scales are different. In all panels, thick(thin) lines correspond to the Newtonian limit(spherical collapse model). Within the framework of the PS and ST formalism, cosmologies with lower values of $\delta_{\star}$ form structure earlier.}
	\label{fig:deltac}
\end{figure*}

\noindent In this linear regime the quintessence field inside the overdensity obeys

\begin{equation}
	\ddot{\phi} + 3\left(H+\frac{\beta}{3}\dot{\delta\phi}-\frac{1}{3}\dot{\delta}_{{\rm m,L}}
\right)\dot{\phi}
				+ \frac{{\rm d}V}{{\rm d}\phi} =-\beta\bar{\rho}_{{\rm m}}(1+\delta_{{\rm m,L}}) \,.
	\label{linearsphrhophinoR}	
\end{equation}

Eqs.~(\ref{sphrhophinoR}-\ref{linearsphrhophinoR}) were first derived in Ref.~\cite{mota} for the case of a variable coupling $\beta(\phi)$. In arriving at this description for the non--linear evolution of a spherical matter perturbation, approximations have been made. Firstly, it is assumed that there is no loss in scalar field energy density from inside the collapsing region~\cite{bruck,mota} -- the quintessence field is assumed to cluster with the CDM and whole system (CDM and scalar field) virializes. It has been argued~\cite{wintergerst} that since the standard spherical collapse model is intrinsically based on gravitational attraction only, it incorrectly describes the evolution of overdensities into the non--linear regime when applied to CQ cosmologies. Indeed, when a dark--dark coupling is introduced, CDM will experience an additional attractive force which is not explicitly included in the spherical collapse formulation. By considering the Newtonian limit of the relativistic non--linear perturbation equations, Ref.~\cite{wintergerst} was able to show that the evolution of the non--linear and linear density contrasts obey

\begin{equation}
	\ddot{\delta}_{{\rm m}}=-(2H+\beta\dot{\bar{\phi}})\dot{\delta}_{{\rm m}}+\frac{4}{3}\frac{\dot{\delta}^{2}_{{\rm m}}}{1+\delta_{{\rm m}}}
										+\frac{1}{2}\bar{\rho}_{{\rm m}}\delta_{{\rm m}}(1+2\beta^{2}) \,,
	\label{PETTnonlin}	
\end{equation}

\noindent and

\begin{equation}
		\ddot{\delta}_{{\rm m,L}}=-(2H+\beta\dot{\bar{\phi}})\dot{\delta}_{{\rm m,L}}+\frac{1}{2}\bar{\rho}_{{\rm m}}\delta_{{\rm m,L}}(1+2\beta^{2}) \,.
	\label{PETTlin}	
\end{equation}

\noindent respectively. We refer the reader to Ref.~\cite{wintergerst} where a detailed comparison to the spherical collapse model is presented. Eq.~(\ref{PETTlin}) is simply the relativistic Eq.~(\ref{CDMgrowthrateCQ}) in standard cosmic time. We immediately notice (as the Newtonian limit demands) that Eqs.~(\ref{PETTnonlin}) and (\ref{PETTlin}) contain no terms involving time derivatives of $\delta\phi$. Also, there is no term analogous to the fourth and third term on the RHS of Eqs.~(\ref{modifiedSphCol}) and (\ref{linearOverdensity}), since to satisfy the Newtonian limit the quintessence field should not contribute to the gravitational potential. Notice that this model reduces to the standard spherical collapse description in the limit of a cosmological constant with zero coupling. As we will now see, this approach can lead to significantly different predictions for $\delta_{\star}(z)$ when compared to the spherical collapse model.

In Fig.~\ref{fig:deltac} we depict the redshift evolution of the parameter $\delta_{\star}(z)$, where the thick lines represent the Newtonian (NWT) prediction and the thin lines the corresponding spherical collapse (SPH) prediction. We depict a range of coupling strengths $\beta>0$. Notice the SPH model prediction of oscillatory behaviour of $\delta_{\star}(z)$ for the uncoupled AS model. This is due to late time oscillations of the field about the minimum of the potential and are progressively damped for stronger couplings (see Section \ref{potentials} for an explanation). These oscillations are not present in the NWT prediction since the quintessence field is assumed not to cluster. In general, for the 2EXP potential (for which perturbation growth is enhanced relative to $\Lambda$CDM for $\beta>0$), the NWT model predicts a higher $\delta_{\star}(z)$ with higher $\beta$, whilst the SPH model predicts lower values of $\delta_{\star}(z)$ with higher $\beta$. For the SUGRA and AS potentials (where perturbation growth is suppressed relative to $\Lambda$CDM for $\beta>0$), the NWT model predicts a lower $\delta_{\star}(z)$ with higher $\beta$, whilst the SPH model predicts higher values of $\delta_{\star}(z)$ with higher $\beta$. For $\beta<0$, this behaviour is reversed. Within the framework of the PS and ST formalism, cosmologies with lower values of $\delta_{\star}$ form structure earlier. \\

In summary, the assumptions made when following the dark energy scalar field into the non–-linear regime, i.e., whether the field is to be treated separately inside and outside an overdense region, greatly effects the critical density at the redshift of collapse. Without proper knowledge of how to treat the non--linear evolution of the field, we compare halo abundance computed using the NWT and SPH approaches. The differences between the SPH and NWT models for some CQ cosmologies are large enough to become manifest in the high--$z$, high--mass tail of the mass function due to the exponential dependence on $\delta_{\star}^{2}$.  It is important however to put these differences into context with the other uncertainties that go into modelling the mass function. As we will discuss in section~\ref{resultsB}, the differences between the NWT and SPH models are secondary compared to the uncertainties associated with the use of the extremal values of $\beta$, obtained by fitting the quintessence models to data. When one then considers the limited accuracy of the fitting functions $f(\sigma,z)$ used in this work, the difference induced in the mass function by using either the SPH or NWT models becomes a secondary concern.

\begin{figure*}[t]
	\begin{tabular}{cc}
		\includegraphics[width=8.7cm]{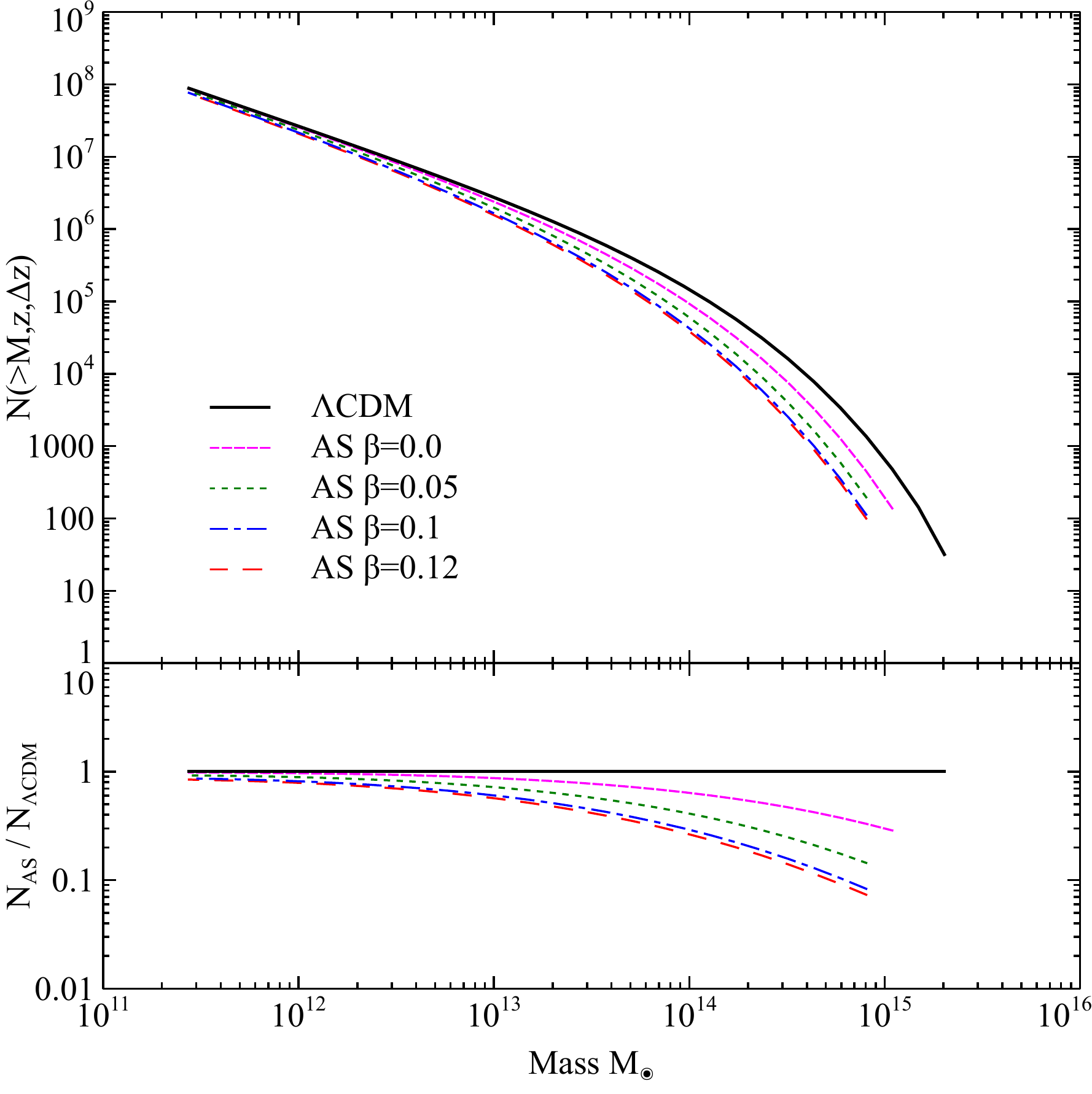} &
		\includegraphics[width=8.7cm]{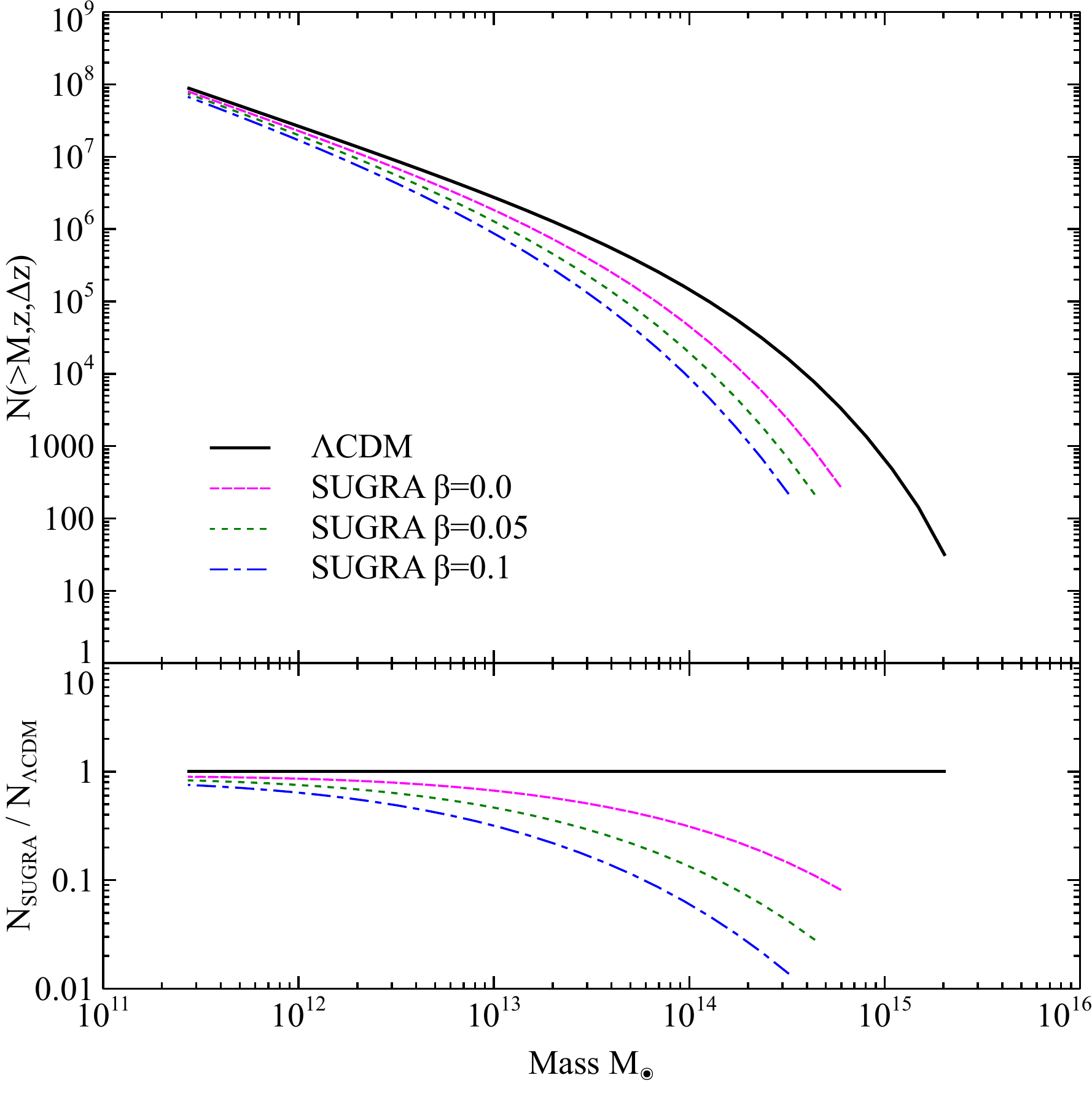} \\	
		\includegraphics[width=8.7cm]{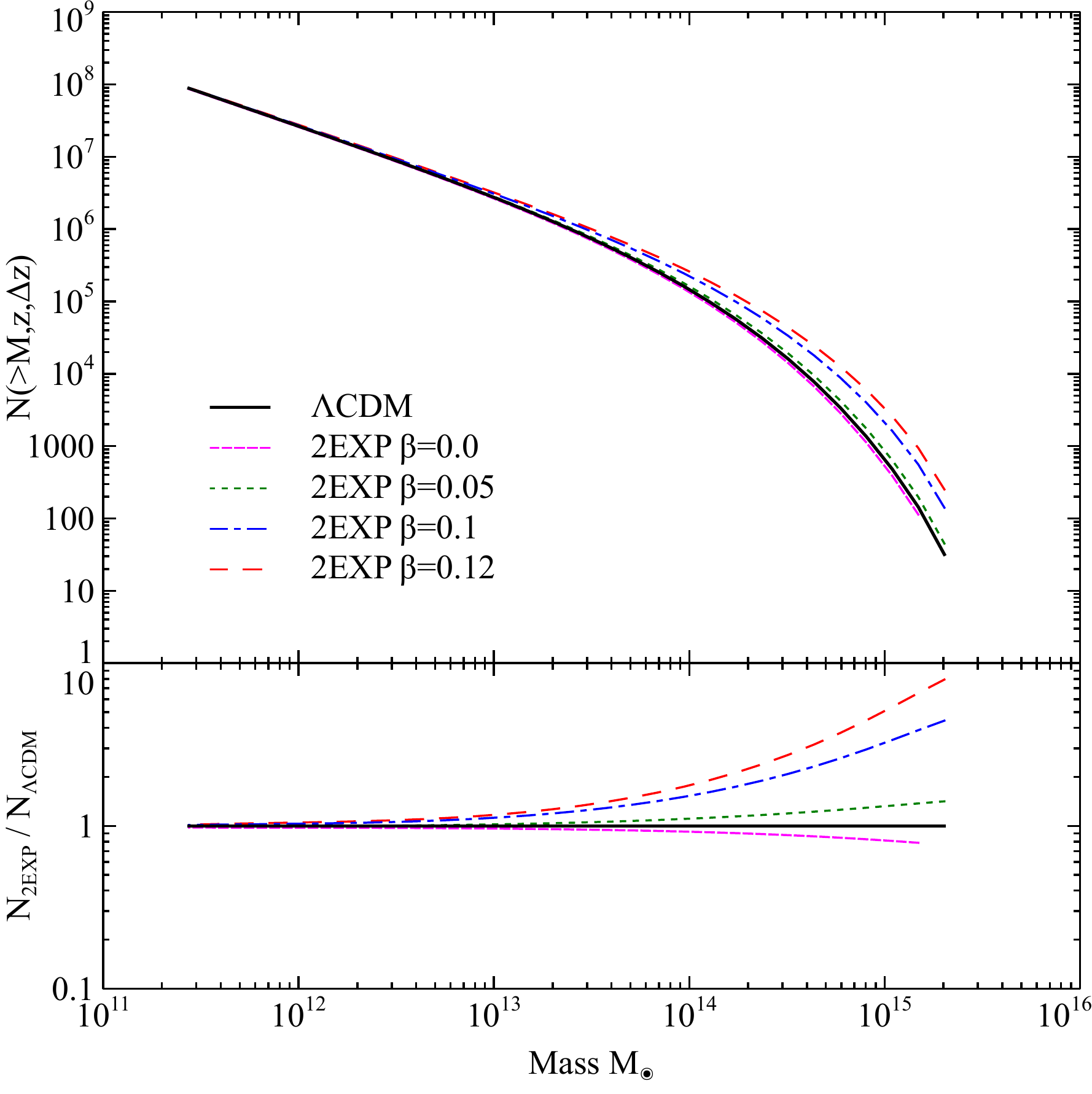}  &
		\includegraphics[width=8.7cm]{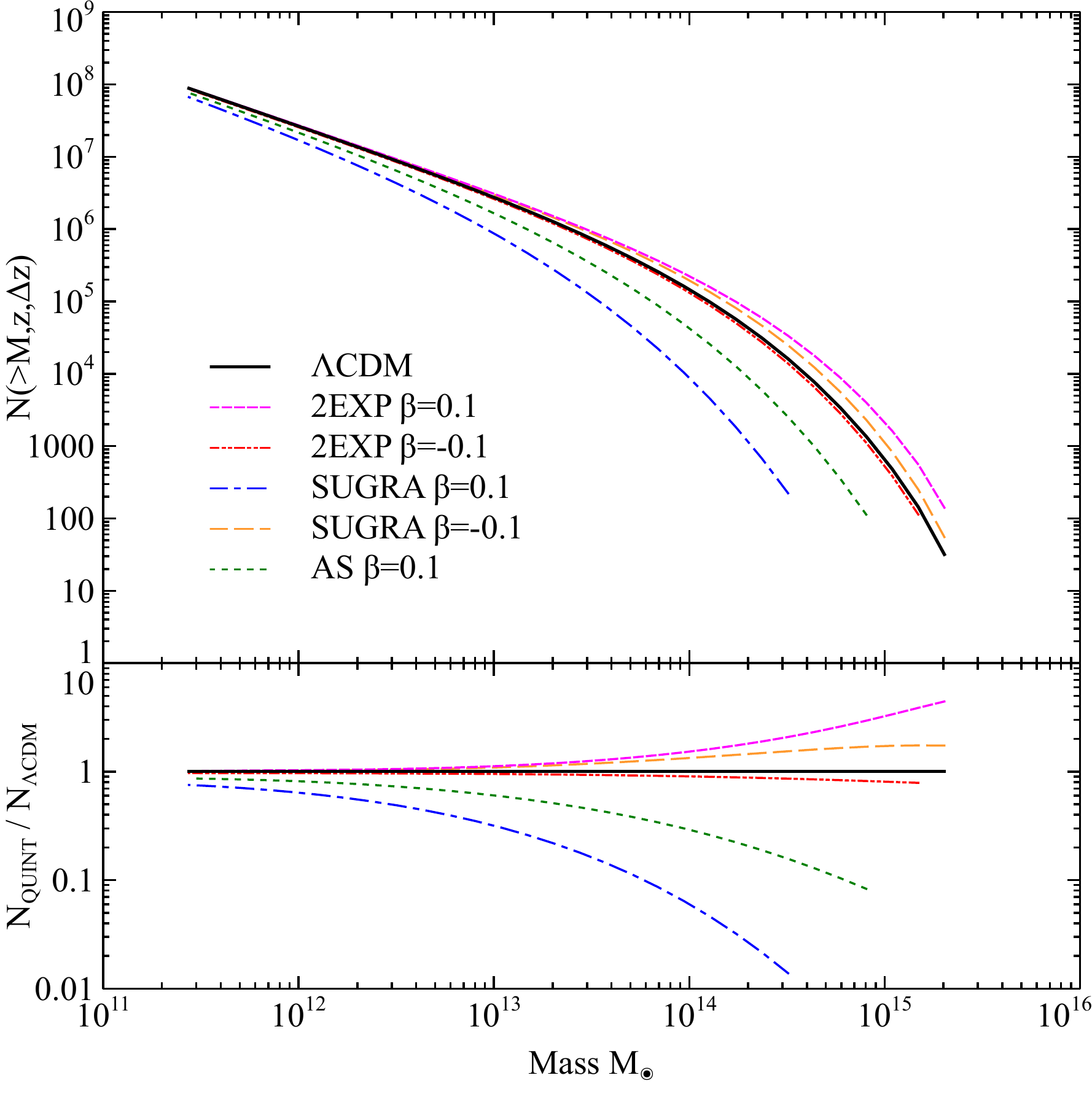}
	\end{tabular}
	\caption{The Jenkins et al. mass function prediction for the mass distribution of CDM haloes with $10^{11}h^{-3}\,\text{M}_{\odot}\leq M\leq 10^{15}h^{-3}\,\text{M}_{\odot}$, within a redshift survey volume $z=0.01-0.4$. \textit{Top left panel}: AS model. \textit{Top right panel}: SUGRA model. \textit{Bottom left panel}: 2EXP model. \textit{Bottom right panel}: A comparison between the three quintessence models for a range of coupling values. Note that the Jenkins et al. mass function does not reach the highest cluster masses, since it is only valid over a limited range of $\sigma$.}
	\label{fig:JK-NvM1}
\end{figure*}

\section{Results}\label{results}

We will now present results for the halo abundance for the coupled quintessence models introduced in Sec.~\ref{potentials} using the mass functions described in Sec.~\ref{massfunction}. To facilitate an easier comparison between models, we begin by normalising each model to the $\Lambda$CDM WMAP7 Maximum Likelihood (ML) cosmological parameter values~\cite{WMAP7}. As the resulting cosmologies are unlikely to be consistent with all observational data, as pointed out in Ref.~\cite{jennings}, we subsequently perform a global fit using the \texttt{CosmoMC} package~\cite{cosmomc}, a Monte Carlo Markov Chain (MCMC) code, to find the best fit values for the cosmological parameters such that each model satisfies the CMB+BAO+SN1a+$H_{0}$ datasets.\\

\begin{figure*}[t]
	\begin{tabular}{cc}
		\includegraphics[width=8.7cm]{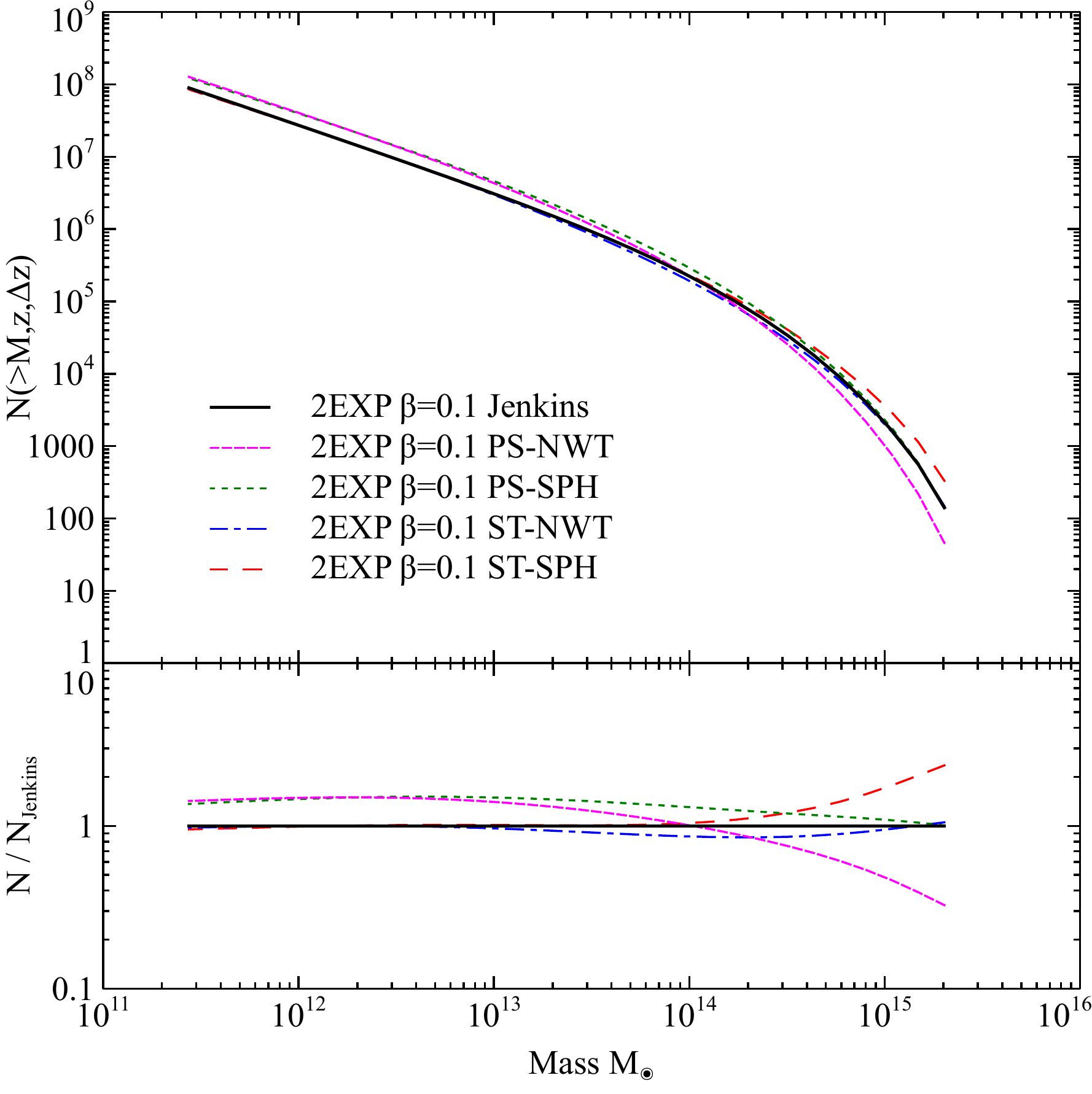} &
		\includegraphics[width=8.7cm]{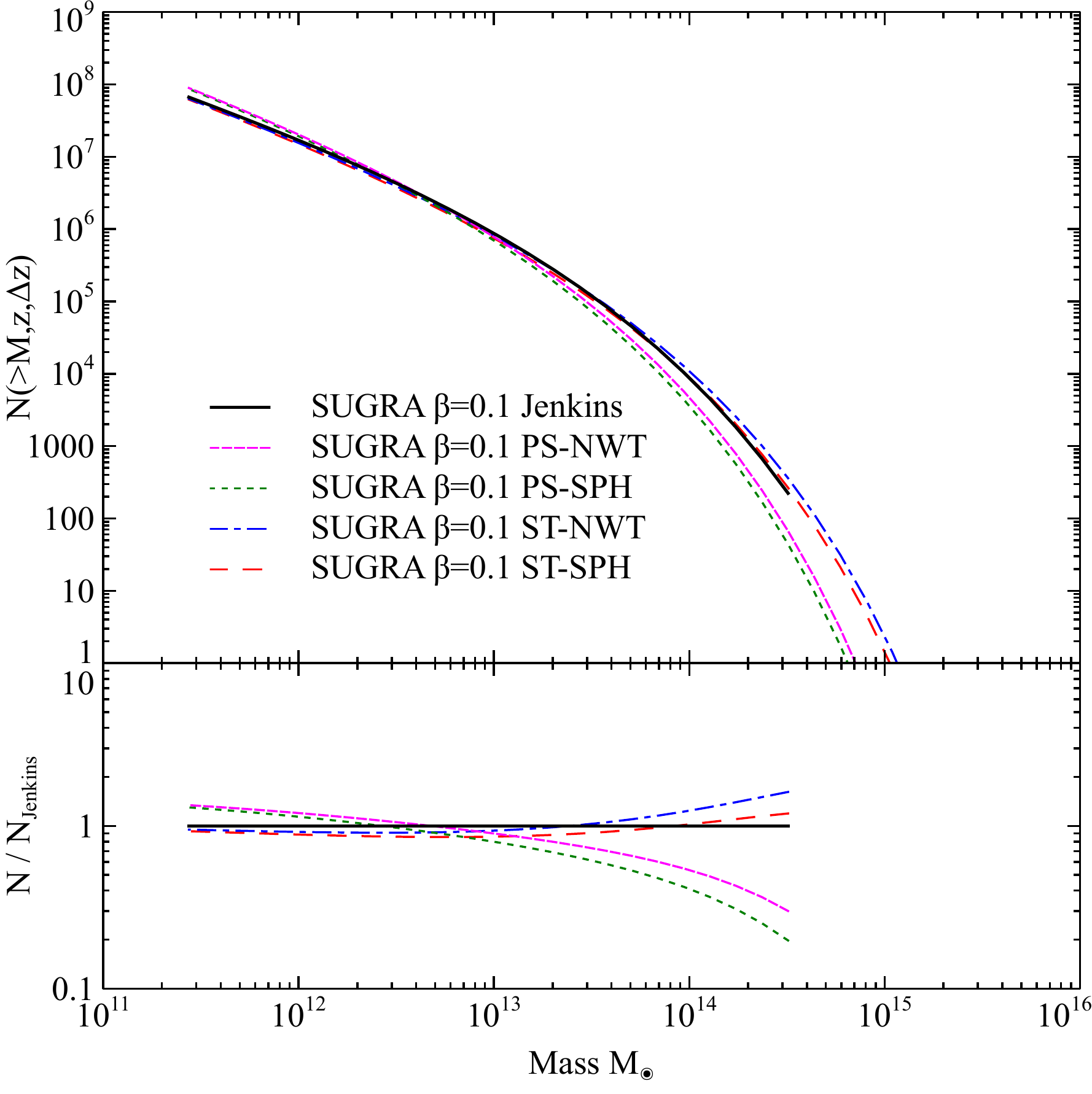}  \\
		\includegraphics[width=8.7cm]{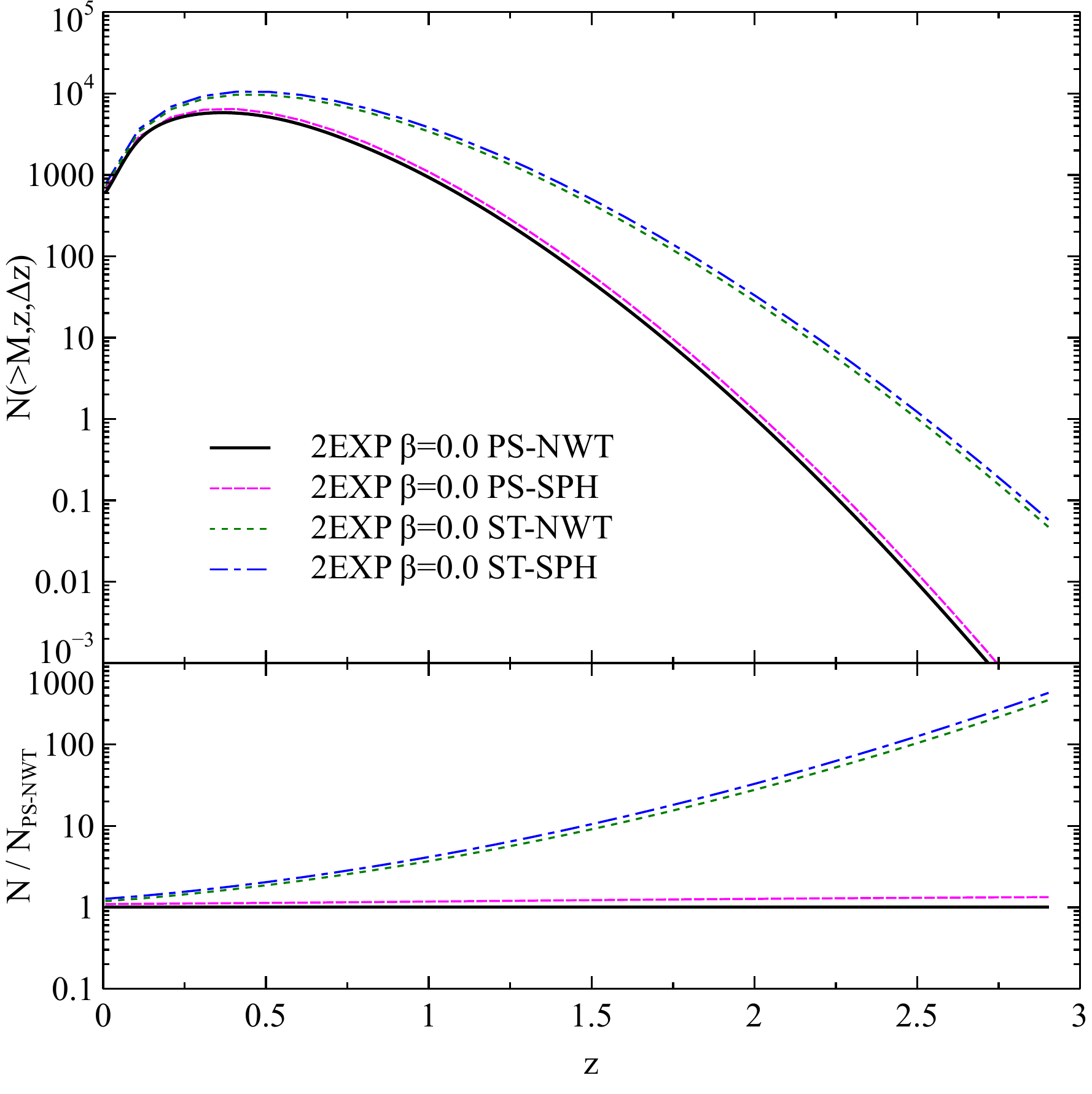} &
		\includegraphics[width=8.7cm]{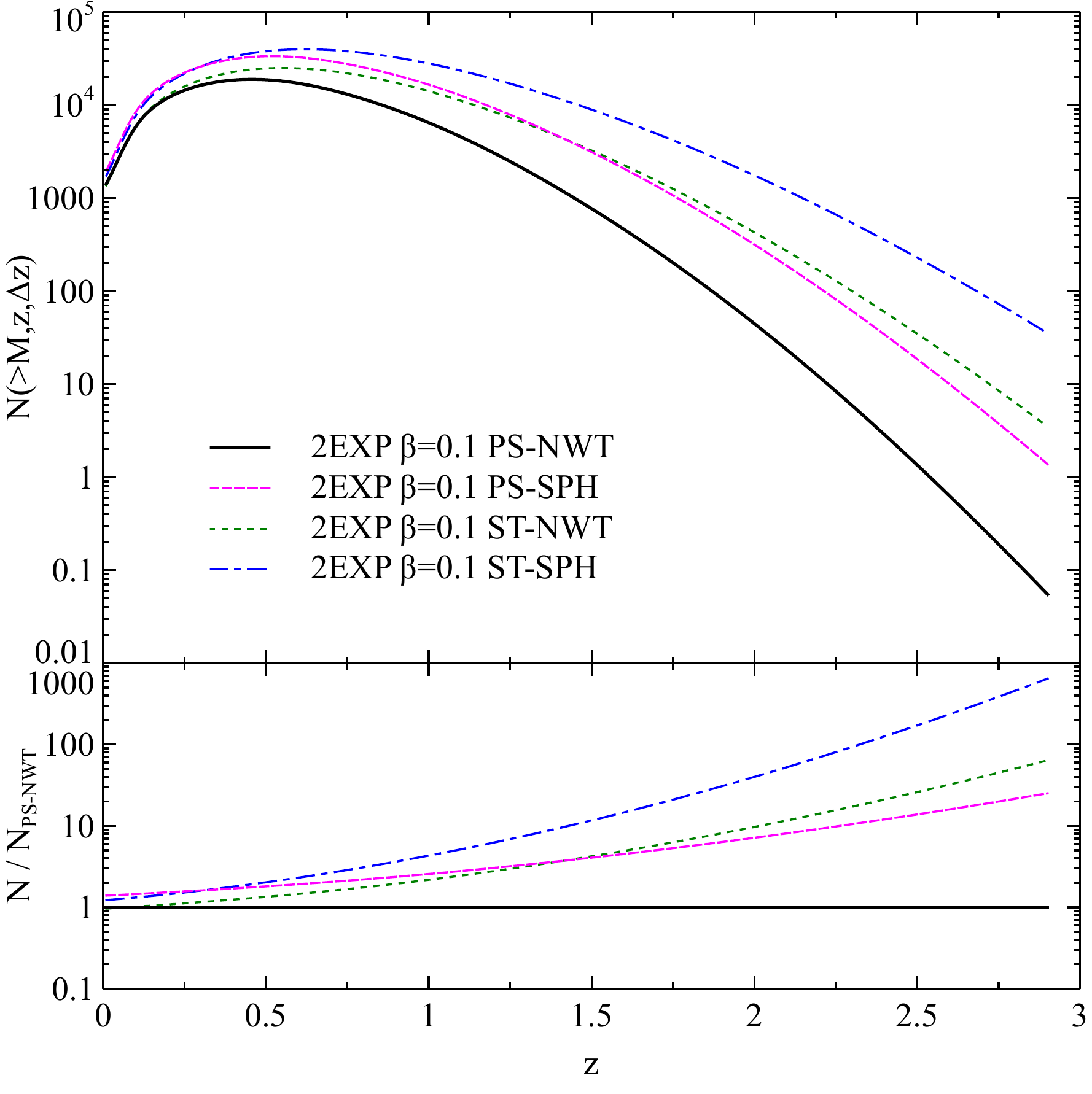}
	\end{tabular}
	\caption{A comparison between the cluster number counts, $\mathcal{N}(>M,z,\Delta z)$, for the different mass functions for $10^{11}h^{-3}\,\text{M}_{\odot}\leq M\leq 10^{15}h^{-3}\,\text{M}_{\odot}$, within a redshift survey volume $z=0.01-0.4$. \textit{Top left panel}: 2EXP model, $\beta=0.1$. \textit{Top right panel}: SUGRA model, $\beta=0.1$. Below each figure the PS and ST predictions are compared to the Jenkins et al. mass function by taking the ratio. Note that the Jenkins et al. mass function does not reach the highest cluster masses, since it is only valid over a limited range of $\sigma$.  \textit{Bottom left panel}: The redshift evolution of the cluster number count, $\mathcal{N}(>M,z,\Delta z)$, for $10^{14}h^{-3}\,\text{M}_{\odot}\leq M\leq 10^{15}h^{-3}\,\text{M}_{\odot}$ in the 2EXP model for $\beta=0$, with redshift bins of  width $\Delta z=0.1$. \textit{Bottom right panel}: 2EXP, $\beta=0.1$. Below each panel, we plot the ratio of halo abundance to that found using the PS--NWT approach.}
	\label{fig:QUINT_JK-PS-ST-NWT-SPH_NvM}
\end{figure*}

\subsection{WMAP $\Lambda$CDM cosmological parameters}
\label{resultsA}

In this subsection we vary the normalisation of the potential and the CDM density, so that all cosmologies share the same $\Lambda$CDM WMAP7 cosmological parameter values~\cite{WMAP7}, given in Table~\ref{mlparamtable} of Appendix~\ref{apdxMLparamValues}. For all uncoupled cosmologies (i.e. $\beta=0$), we find a lower abundance of high mass haloes compared to $\Lambda$CDM, when normalised to CMB fluctuations. This is in agreement with previous studies (c.f.~\cite{mortonson} and references therein). 

In Fig.~\ref{fig:JK-NvM1} we show the theoretically expected abundance of CDM haloes, $\mathcal{N}(>M,z,\Delta z)$, in the mass range $10^{11}\,\text{M}_{\odot}h^{-3}\leq M\leq 10^{15}h^{-3}\,\text{M}_{\odot}$ and in a redshift volume $z=0.01-0.4$, using the Jenkins et al. mass function. This mass range extends from typical galaxy size halos, up to galaxy cluster masses. Below each figure we plot the ratio of the quintessence cosmology to $\Lambda$CDM to facilitate an easier comparison.  Starting with the AS model, the Jenkins et al. mass function predicts a reduction in cluster abundance in the high mass tail. Numbers of the most massive clusters are further reduced when a coupling $\beta>0$ is introduced. The same trend is true of the SUGRA models, however suppression in the high mass tail is more severe, with differences compared to $\Lambda$CDM reaching almost two orders of magnitude for haloes of mass $M>10^{14}h^{-3}\,\text{M}_{\odot}$ for $\beta=0.1$. Here, the sign of the coupling parameter and the shape of the effective potential in which the field evolves dictates that there is a continual flow of energy from the quintessence field to the CDM, i.e., $Q_{({\rm c})\,0}<0$. Considering the arguments of Sections \ref{potentials} and \ref{perturbations}, the effect of this direction of energy exchange is to suppress structure formation relative to $\Lambda$CDM, regardless of the fifth force mediated by the quintessence field. 

As one moves from $\Lambda$CDM to the uncoupled 2EXP model, we expect to see a reduction in the number of high mass ($M>10^{14}h^{-3}\,\text{M}_{\odot}$) clusters. When a weak positive coupling ($\beta=0.05$) is introduced, $\Lambda$CDM abundances are recovered, and as the coupling strength is increased we expect to see an excess of high mass clusters, since $Q_{({\rm c})\,0}>0$. In the SUGRA model, for increasingly more negative couplings, $Q_{({\rm c})\,0}>0$ and $\Lambda$CDM abundances are recovered and then surpassed. In the 2EXP model, for weak negative couplings $\beta\sim-0.05$, the number of high mass clusters is suppressed relative to $\beta=0$. For stronger, more negative couplings, high mass cluster numbers in the uncoupled model are recovered and with $\beta<\sim-1.2$, $\Lambda$CDM numbers are surpassed. These effects are  demonstrated in the bottom--right panel of Fig.~\ref{fig:JK-NvM1} where we compare the three potentials for selected positive and negative coupling values. As has been noticed previously~\cite{mainini}, the differences between the uncoupled quintessence models and $\Lambda$CDM can be partially or even totally reduced, (i.e. they become degenerate), by introducing a weak DE--CDM coupling.

\begin{figure*}[t]
	\begin{tabular}{cc}
		\includegraphics[width=9cm]{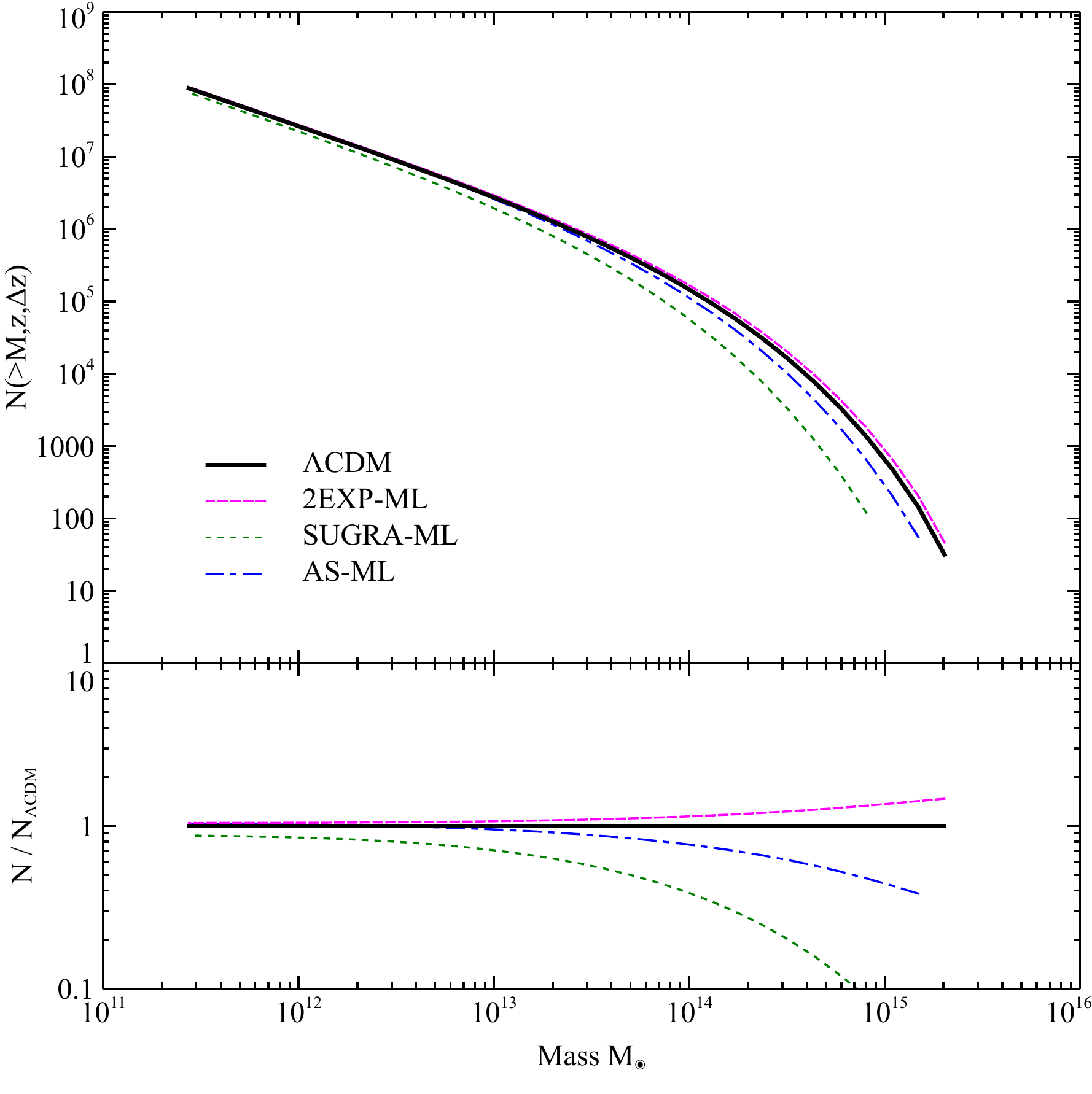} &
		\includegraphics[width=9cm]{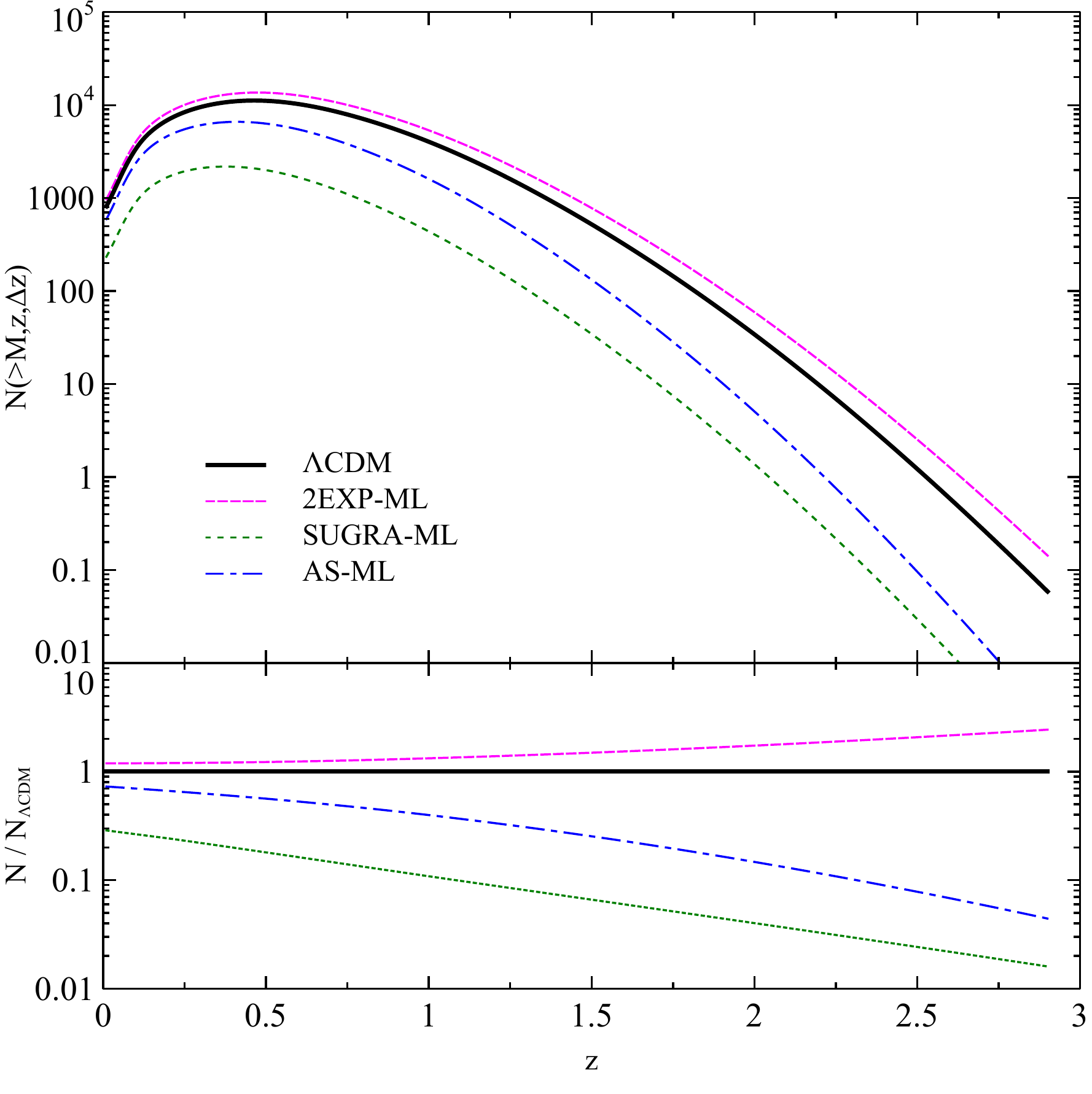} 
	\end{tabular}
	\caption{\textit{Left panel}: The number counts of CDM haloes with $10^{11}h_{\Lambda}^{-3}\,\text{M}_{\odot}\leq M\leq 10^{15}h_{\Lambda}^{-3}\,\text{M}_{\odot}$, within a redshift survey volume $z=0.01-0.4$ found using the Jenkins et al. mass function. \textit{Right panel}: The redshift evolution of the cluster number count, $\mathcal{N}(>M,z,\Delta z)$, for $10^{14}h_{\Lambda}^{-3}\,\text{M}_{\odot}\leq M\leq 10^{15}h_{\Lambda}^{-3}\,\text{M}_{\odot}$ using the ST--NWT mass function. $h_{\Lambda}$ is the value of $h$ in the $\Lambda$CDM cosmology. We use bins of redshift width $\Delta z=0.1$.}
	\label{fig:QUINT-ML}
\end{figure*}

In the upper panels of Fig.~\ref{fig:QUINT_JK-PS-ST-NWT-SPH_NvM} we compare the halo abundances calculated using the PS and ST mass functions to the Jenkins et al. mass function, for the 2EXP and SUGRA models with $\beta=0.1$. We use both the NWT and SPH methods for calculating $\delta_{\star}(z)$. Starting with the 2EXP cosmology, ST--NWT agrees well with the Jenkins prediction, whilst ST--SPH predicts a slight excess of high mass haloes compared to both Jenkins and ST--NWT. The PS mass function predicts a slightly greater abundance of low mass clusters compared to Jenkins. At the high mass end, the PS--SPH model lies above the Jenkins prediction, whilst the PS--NWT predicts a reduction in halo abundance. The same general trend between the PS, ST and Jenkins et al. mass functions is true of the SUGRA cosmology, however this time the spherical collapse model predicts a lower number of haloes compared to the Newtonian model. These differences are because in the 2EXP cosmology, the NWT model predicts a higher  $\delta_{\star}(z)$ with higher $\beta$, whilst the SPH model predicts lower values of $\delta_{\star}(z)$ with higher $\beta$. Within the framework of the PS and ST formalism, cosmologies with lower values of $\delta_{\star}$ form structure earlier, and hence the SPH model predicts a greater abundance of haloes. The same argument applies to the SUGRA cosmology, where for $\beta=0.1$, the NWT model predicts lower values of $\delta_{\star}(z)$ compared to the SPH model. For the AS cosmology, we found that the differences between the NWT and SPH model predictions for $\delta_{\star}(z)$ were negligible when incorporated into the PS and ST mass functions.

Focusing on the high mass tail of the mass function, since this is where deviations from $\Lambda$CDM are most prominent, we found that for stronger coupling strengths the 2EXP cosmology has a higher abundance of massive objects at high redshift compared to $\Lambda$CDM (for a given mass function), whilst deviations from $\Lambda$CDM remain much less than one order of magnitude at low $z<0.5$ redshifts. The uncoupled and coupled SUGRA and AS cosmologies have fewer massive objects at high and low redshift and this reduction is enhanced for stronger couplings. In the lower panels of Fig.~\ref{fig:QUINT_JK-PS-ST-NWT-SPH_NvM} we show the redshift evolution of the cluster number count for objects of mass $10^{14}h^{-3}\,\text{M}_{\odot}\leq M\leq 10^{15}h^{-3}\,\text{M}_{\odot}$ in the 2EXP model. For the uncoupled model (left panel), the NWT and SPH predictions for $\delta_{\star}$ agree well over the redshift range of interest, with PS theory underestimating the number of high redshift objects compared to the ST mass function. For the $\beta=0.1$ model (right panel) the different redshift evolution of $\delta_{\star}(z)$ predicted by the NWT and SPH models is clearly manifest at high redshifts. Similar, non--negligible differences occur between the NWT and SPH models in the SUGRA cosmologies.\\


\subsection{Maximum Likelihood cosmological parameters}\label{resultsB}

In this subsection we present mass function results for the 2EXP, SUGRA and AS models using the ML cosmological parameter values obtained from a MCMC global fit. 

We take the set of base parameters (see Appendix~\ref{apdxMLparamValues} for an explanation of these parameters) 

\begin{equation}
		\boldsymbol\vartheta  = \{ \Omega_{\rm b}h^{2},\, \Omega_{\rm c}h^{2},\, H_0\,, \tau,\, \log{[10^{10}\Delta^{2}_{\mathcal{R}}(k_{0})]},\, n_{\rm s},\, \beta \} \,,
	\label{paramSet}	
\end{equation}

\noindent and use a modified version of the \texttt{CosmoMC} package~\cite{cosmomc} to sample from the joint posterior distribution of the parameters,

\begin{equation}
		\mathcal{P}(\boldsymbol\vartheta | \boldsymbol x) = \frac{\mathcal{L}(\boldsymbol x | \boldsymbol\vartheta)\mathcal{P}(\boldsymbol\vartheta)} 
			{\int {\rm d}\boldsymbol\vartheta\, \mathcal{L}(\boldsymbol x | \boldsymbol\vartheta)\mathcal{P}(\boldsymbol\vartheta)} \,,
	\label{jointPosterior}	
\end{equation}

\noindent where $\mathcal{L}(\boldsymbol x | \boldsymbol\vartheta)$ is the likelihood of the data $\boldsymbol x$ given the model parameters $\boldsymbol\vartheta$ and $\mathcal{P}(\boldsymbol\vartheta)$ is the prior probability density.

We use the CMB+BAO+SN1a+$H_{0}$ datasets (see Appendix~\ref{apdxMLparamValues}) and apply a Gelman--Rubin~\cite{gelman} convergence criterion of $R-1\lesssim0.03$ across eight chains for each model. We assume purely adiabatic initial conditions with no tensor contribution.  We do not vary the parameters $\lambda_{n}$ in the quintessence potentials; they remain fixed with the values given in Sec.~\ref{potentials}. We take this approach as our goal is to investigate the observability of couplings of order $\beta \sim 0.1$. Finding a best fit including varying the potential parameters is therefore not necessary for this particular approach although of course there is nothing to stop them being included in general. This approach also allows us to compare our results with those from Sec.~\ref{resultsA}, and assess the effect of (erroneously) using the $\Lambda$CDM cosmological parameters. For the 2EXP and SUGRA models we consider couplings in the range $\beta\,\in\,[-0.4,\,0.4]$  i.e. including negative and zero couplings, whilst for the AS model, $\beta\,\in\,[0.0,\,0.4]$, since it is not possible to realise a cosmology with $\Omega_{\phi}\sim0.7$ today for $\beta \leq -0.05$. We assume flat priors ($\Omega_{{\rm K}}=0$) on all base parameters. In Table~\ref{mlparamtable} of Appendix~\ref{apdxMLparamValues} we list the ML values of this parameter set and the derived parameters for $\Lambda$CDM, 2EXP, SUGRA and AS. Also shown in brackets are  the extremal values of the parameters in the full $n$--dimensional 68\% confidence region. 

\begin{figure}[t]
\includegraphics[width=8.5cm]{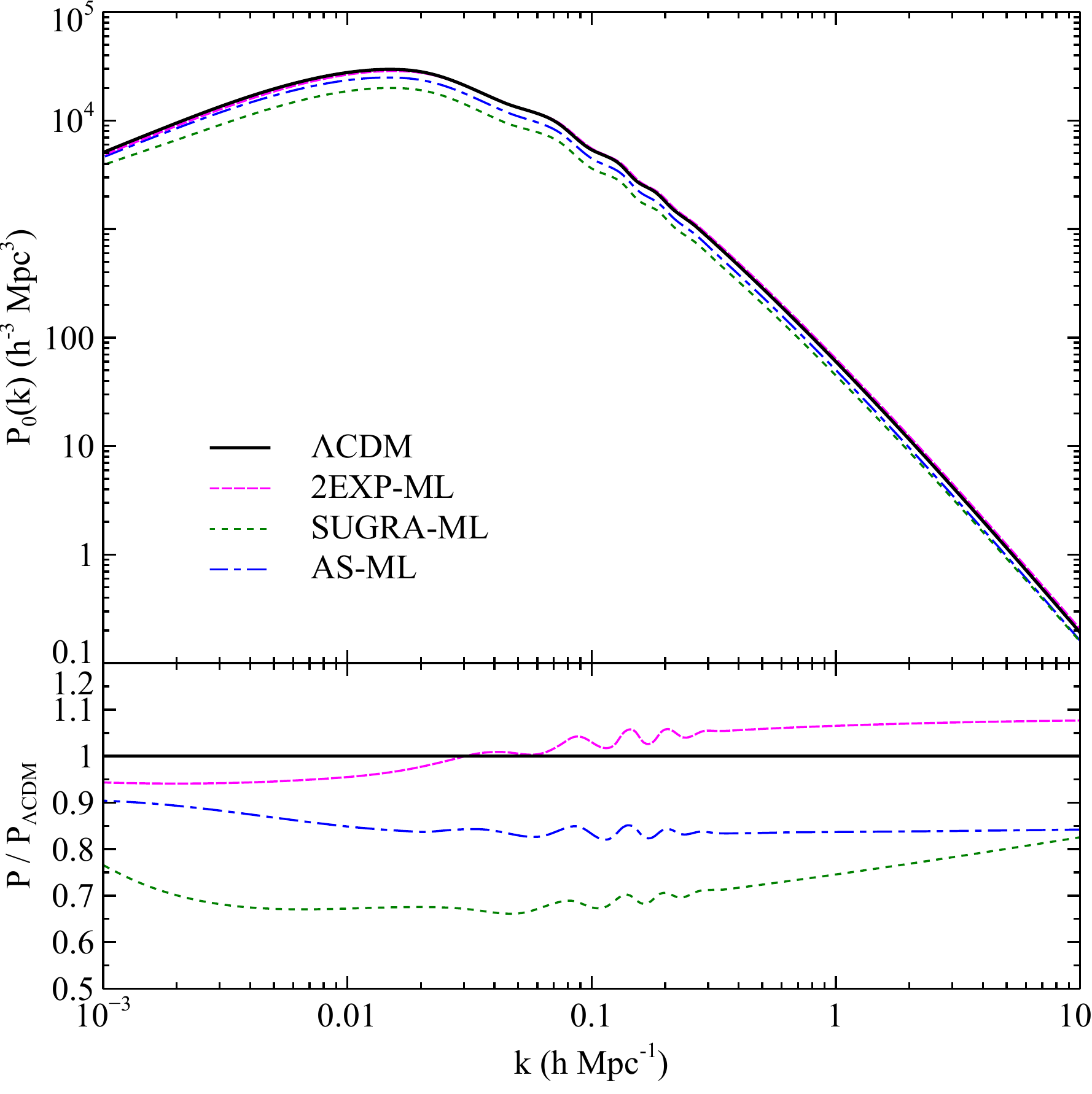}
\caption{The linear total matter ($\rho_{{\rm c}}+\rho_{{\rm b}}$) power spectrum at $z=0$, $P_{0}(k)$, for the three quintessence cosmologies with ML cosmological parameter values. Note that all spectra have been normalised to CMB fluctuations, which are on larger scales than included in the figure.}
\label{fig:QUINT-ML_powerSpectrum}
\end{figure}

In Fig.~\ref{fig:QUINT-ML}, using the parameter values in Table~\ref{mlparamtable} of Appendix~\ref{apdxMLparamValues}, we show halo abundance as a function of mass (left panel) and as a function of redshift (right panel).  The positive value of the coupling parameter ($\beta=0.033$) for the 2EXP--ML cosmology acts to slightly enhance the formation of the most massive haloes, which combined with the fractionally slower Hubble expansion rate ($H_{0}=69.86\,{\rm km\, s}^{-1} \, {\rm Mpc}^{-1}$) and higher matter density ($\Omega_{{\rm m}}=0.273$) mean that more massive objects have time to form during the matter era in the 2EXP--ML model relative to $\Lambda$CDM. These differences do not exceed a factor of $\sim2$ across the entire mass and redshift range of interest. Although the fractionally slower Hubble expansion rate ($H_{0}=69.5\, {\rm km\, s}^{-1} \, {\rm Mpc}^{-1} $) for the SUGRA--ML cosmology will act to enhance structure formation, the lower matter density fraction ($\Omega_{{\rm m}}=0.221$), negative coupling parameter ($\beta=-0.078$) and suppressed linear power spectrum amplitude across all relevant scales (see Fig.~\ref{fig:QUINT-ML_powerSpectrum}) mean that less massive objects have time to form during the matter era in the SUGRA--ML model relative to $\Lambda$CDM, with abundances further decreasing with redshift, reaching one order of magnitude difference by $z=1.5$. The slightly higher matter density of the AS--ML model ($\Omega_{{\rm m}}=0.273$) compared to $\Lambda$CDM slightly raises the halo abundance, however the suppressed power spectrum amplitude (see Fig.~\ref{fig:QUINT-ML_powerSpectrum}) still means that numbers of the most massive high--$z$ clusters still fall short of $\Lambda$CDM predictions. Comparing these results with those from Section \ref{resultsA}, (where each model shared the same cosmological parameter values as $\Lambda$CDM) we see that the departures from $\Lambda$CDM halo abundance are not as great in the ML case. As the CQ models are by construction such that they are now consistent with a wider range of observational data, this is not unexpected. 

The differences in the threshold of collapse, $\delta_{\star}$, obtained from the spherical collapse model and from the Newtonian limit of the non--linear perturbation equations for the AS--ML model had a negligible effect on the halo counts when incorporated into the mass function. We found non--negligible differences between the two approaches when used for the SUGRA and 2EXP--ML models, where the lower value of $\delta_{\star}$ obtained from the spherical collapse model leads to higher halo abundance relative to the Newtonian prediction. As both the spherical collapse model and the model of Ref.~\cite{wintergerst} are based on approximations, we can, to some extent, view these differences as another source of uncertainty when making predictions for galaxy and cluster number counts. This uncertainly must be put into context however, alongside the other uncertainties, such as parameter constraints on $\beta$ and the accuracy of the semi--analytic form of the mass function. For example, for all coupled quintessence models, we find the differences induced in the mass function by taking the $1\sigma$ $n$--dimensional extremal values of the coupling parameter $\beta$ given in Table~\ref{mlparamtable} of Appendix~\ref{apdxMLparamValues}, to be larger than the differences induced in the mass function by using either the NWT or SPH models. When one then considers the limited accuracy of the fitting functions, $f(\sigma,z)$, used in this work, the modelling of the halo collapse threshold becomes a secondary concern. 


\section{Conclusions}\label{conclusions}

We have studied the abundance of CDM haloes in three different coupled quintessence cosmologies, using three common analytic forms for the halo mass function. For all uncoupled cosmologies, we found a lower abundance of high mass haloes compared to $\Lambda$CDM, when normalised to CMB fluctuations, which is in agreement with previous studies (cf.~\cite{mortonson} and references therein). 

When the quintessence field is coupled to CDM through a coupling given by Eq.~(\ref{defBeta}) the direction of energy transfer is determined by the sign of the coupling parameter and on the effective potential in which the field evolves. For the models considered in this work, if there is a continual flow of energy from the quintessence scalar field to the CDM component, then the predicted number of CDM haloes can only lie below that of the uncoupled model (and therefore lower than $\Lambda$CDM) when each model shares the same cosmological parameters today. If the transfer of energy is from the CDM component to the quintessence field, then the predicted number of CDM haloes can be above or below $\Lambda$CDM depending on the strength of the coupling. Ensuring that the models are consistent with observational data by conducting a MCMC maximum likelihood analysis, the SUGRA--ML and AS--ML models predict reduced numbers of massive objects at high redshift relative to $\Lambda$CDM, whilst the 2EXP--ML model yields an excess of the most massive haloes, increasing with redshift. Since the cosmological parameter values obtained from the maximum likelihood analysis are different for each model, the predicted halo number counts for each model can differ from the counts obtained when each model shares the same cosmological parameter values today, for the same coupling value $\beta$. This reinforces the point made in \cite{jennings} that it is important to work with observationally consistent cosmological models when making predictions for cluster number counts, since the CDM halo mass function is exponentially sensitive to parameters such as $\Omega_{{\rm m}}$ .

It is intriguing that a coupled quintessence model can provide a cosmology that is consistent with observational data and produce an excess of high mass haloes. In light of the recent detections of massive high--$z$ galaxy clusters~\cite{bremer, mancone, muchovej}, a weakly coupled 2EXP quintessence model may help to alleviate tensions between these observations and the standard $\Lambda$CDM paradigm, for which these objects should be extremely rare. This also suggests that CQ models where the coupling function is time dependent, $\beta(t)$, may serve as a viable alternative to $\Lambda$CDM (see also~\cite{baldib}). 

We have also shown that the assumptions made when modelling halo collapse in coupled dark energy cosmologies (NWT or SPH), reflects strongly in the high--$z$, high--mass tail of the mass function due to the exponential dependence on the threshold of collapse parameter $\delta_{\star}^{2}$. When this uncertainty is put into context however, alongside the other uncertainties such as parameter constraints on $\beta$ and the accuracy of the semi--analytic form of the mass function, the modelling of the halo collapse conditions becomes a secondary concern. That said, recent advances in mass function theory~\cite{pierStefano,Maggiore2009rv}, have allowed for a more robust calculation of the mass function. This formalism however still relies on a model for the halo collapse conditions. If it were applied to coupled quintessence, the approximations made when calculating the threshold of collapse, $\delta_\star(z)$, would become an important contributor to the total error budget. It will become important to be able to model halo collapse correctly in order to make best use of this new approach. With the arrival of greatly improved high redshift cluster surveys such as the Dark Energy Survey (DES), \textit{Planck} and eventually LSST, the mass distribution of the most massive objects in the Universe will act as a powerful tool which may be used to constrain coupled quintessence models.


\begin{acknowledgments}
We would like to thank Timothy Clemson, Anthony Brookfield and Costas Skordis for useful discussions and for help with numerical work. ERMT is supported by the University of Nottingham. EJC and AMG are supported by STFC. EJC acknowledges the Royal Society for financial support. 
\end{acknowledgments}


\begin{center}
\begin{table*}[ht]
\hfill{}
\begin{tabular}{c|c|c|c|c|c}
\hline
\hline
Class   &   Parameter          			&   $\Lambda$CDM WMAP7--ML  &      2EXP--ML  (68\%-full)    &       SUGRA--ML (68\%-full)     &   AS--ML   (68\%-full)	\\
\hline
Primary & $100\Omega_{\rm b} h^2$  	&   2.227     		                 &   2.244 (2.124 -- 2.392)	   &	2.329 (2.190 -- 2.482)         &  2.281 (2.163 -- 2.442)	\\
 & $\Omega_{{\rm c}} h^2$  			&   0.111			                  &   0.111 (0.101 -- 0.121)	   & 	0.083 (0.076 -- 0.910)    	&  0.116 (0.108 -- 0.127)     \\
& $H_0$  (${\rm km\, s}^{-1} \, {\rm Mpc}^{-1}$)   & 71.4    	                  & 69.86  (65.86 -- 74.18)     &      69.50	 (65.40 -- 74.35)	&   71.4 (67.20 -- 74.92) \\
 & $\tau$            				&   0.086					&   0.089 (0.052 -- 0.131)	   &	0.091  (0.055 -- 0.140)	&  0.093 (0.055 -- 0.134)  \\ 
 & ${\rm log}[10^{10}\Delta^{2}_{\mathcal{R}}(k_{0})]^{\dagger}$	
                                                               & 3.170                                         &   3.08 (2.99 -- 3.18)	            &	3.07   (2.99 -- 3.18)       	&  3.07   (2.98 -- 3.19)\\
 & $n_s$             		  			&	0.969				&   0.967 (0.941 -- 1.01)         &	 1.013 (0.983 -- 1.049)	&  0.977 (0.946 -- 1.010)  \\
 & $\beta$       			     		&	--					&  0.033  (-0.169 -- 0.096)    &	-0.078 (-0.123 -- -0.044)	&  0.0048 (0.0 -- 0.046) \\
 & 							&						&					&						&  \\
\hline
Derived & $\Omega_\Lambda$  		&      0.738					&   0.727 (0.675 -- 0.768)	& 	0.779 (0.740 -- 0.815)	&  0.727 (0.677 -- 0.759) \\
 & $t_{0}$  (Gyr)     	 			&	13.71			         &   13.74 (13.37 -- 14.09)	&	13.69 (13.24 -- 14.0)		&  13.52 (13.22 -- 13.89) \\
 & $\Omega_m$        				& 	0.262	 			&   0.273 (0.231 -- 0.325) 	&	0.221 (0.184 -- 0.259)	&  0.273 (0.240 -- 0.323)\\
 & $z_{re}$          	 				&	10.3					&   10.7   (7.4 -- 13.9) 	&   	10.5	(7.4 -- 13.7)		&  11.0 (7.7 -- 14.2)
\end{tabular}
\hfill{}
\caption{Maximum Likelihood values for $\Lambda$CDM and the three coupled quintessence models. For the coupled quintessence we also show in brackets the extremal values of the parameters in the full $n$--dimensional 68\% confidence region. $^{(\dagger)}$ $\Delta^{2}_{\mathcal{R}} (k)=k^{3}\mathcal{P}_{\mathcal{R}} (k)/(2\pi^2)$. For $\Lambda$CDM WMAP7-ML, $k_0=0.002\,\mathrm{Mpc}^{-1}$. For 2EXP-ML, SUGRA-ML and  AS-ML, $k_0=0.05\,\mathrm{Mpc}^{-1}$.}
\label{mlparamtable}
\end{table*}
\end{center}

\appendix

\section{Maximum Likelihood parameter values}\label{apdxMLparamValues}

Table \ref{mlparamtable} gives the Maximum Likelihood (ML) cosmological parameter values for each coupled quintessence cosmology, obtained by performing a global fit using the \texttt{CosmoMC} package~\cite{cosmomc}, a Monte Carlo Markov Chain (MCMC) code. We employed a Gelman--Rubin~\cite{gelman} convergence criterion of $R-1\lesssim0.03$ across eight chains for each model. We assumed flat priors ($\Omega_{{\rm K}}=0$) on all base parameters:

\begin{equation}
		\boldsymbol\vartheta  = \{ \Omega_{{\rm b}}h^{2},\, \Omega_{{\rm c}}h^{2},\, H_0,\, \tau,\, \text{log}[10^{10}\Delta^{2}_{\mathcal{R}}(k_{0})],\, n_{{\rm s}},\, \beta \} \,,
	\label{paramSetAppdx}	
\end{equation}

\noindent and we assumed purely adiabatic initial conditions with no tensor contribution. Here, $\Omega_{{\rm b}}h^{2}$ and $\Omega_{{\rm c}}h^{2}$ are the physical baryon and cold dark matter densities respectively, $\tau$ is the reionization optical depth, $\Delta^{2}_{\mathcal{R}}(k_{0})$ is the primordial superhorizon power in the curvature perturbation on scales $k_{0}\,\text{Mpc}^{-1}$, $n_{{\rm s}}$ is the scale spectral index and $\beta$ is the DE--CDM coupling parameter. For the 2EXP and SUGRA models, we consider couplings in the range $\beta\,\in\,[-0.4,\,0.4]$, whilst for the AS model, $\beta\,\in\,[0.0,\,0.4]$, since it is not possible to realise a cosmology with $\Omega_{\phi}\sim0.7$ today for $\beta<\sim-0.05$. The parameters $\lambda_{n}$ in the quintessence potentials were kept fixed with the values given in Sec.~\ref{potentials}.

The observational constraints we use to evaluate the ML parameter values for each coupled quintessence cosmology are local distance measures of the Hubble constant ($H_{0}$) and Type Ia Supernovae (SNIa) and precision measurements of the Cosmic Microwave Background (CMB) and the Baryon Acoustic Oscillations (BAO). For the CMB, we use the 7--year data release from the WMAP satellite (WMAP7)~\cite{WMAP7data} and utilise the likelihood code available at the LAMBDA website~\cite{lambdahttp}. We compute the CMB angular power spectra using a modified CAMB code, described in section \ref{potentials}. The Hubble constant measurements are taken from the SHOES team~\cite{SHOES}, based on geometric distance measurements of SN1a at  $0.023< z< 0.1$. We use the BAO constraints found in Ref.~\cite{BAOsloan} and finally, we use the SNIa sample of Ref.~\cite{kessler}.


\end{document}